\documentclass[manuscript,screen]{acmart}
\usepackage{comment}
\usepackage{lipsum}
\usepackage{booktabs}
\usepackage{tabularx}
\usepackage{graphicx}
\usepackage{longtable}
\usepackage{tabu}

\newcommand{\todo}[1]{\textcolor{red}{\textbf{TODO{: #1}} }} 

\AtBeginDocument{%
  \providecommand\BibTeX{{%
    \normalfont B\kern-0.5em{\scshape i\kern-0.25em b}\kern-0.8em\TeX}}}

\setcopyright{acmcopyright}
\copyrightyear{2022}
\acmYear{2022}
\acmDOI{10.1145/1122445.1122456}

\acmConference[ACM CSUR]{ACM CSUR}{2023}{Online}
\acmBooktitle{ACM CSUR}
\acmPrice{15.00}
\acmISBN{978-1-4503-XXXX-X/18/06}

\begin{document}

\title[From Digital Media to Empathic Reality]{From Digital Media to Empathic Reality: A Systematic Review of Empathy Research in Extended Reality Environments}

\author{Ville Paananen}
\email{ville.paananen@oulu.fi}
\orcid{0000-0001-7377-9770}
\affiliation{
    \institution{University of Oulu}
    \city{Oulu}
    \country{Finland}
}

\author{Mohammad Sina Kiarostami}
\email{Mohammad.Kiarostami@oulu.fi}
\orcid{0000-0002-6100-6567}
\affiliation{
    \institution{University of Oulu}
    \city{Oulu}
    \country{Finland}
}

\author{Lik-Hang Lee}
\email{likhang.lee@kaist.ac.kr}
\orcid{0000-0003-1361-1612}
\affiliation{
    \institution{KAIST, Korean Advanced Institute of Science and Technology}
    \city{Daejeon}
    \country{South Korea}
}

\author{Tristan Braud}
\orcid{0000-0002-9571-0544}
\email{braudt@ust.hk}
\affiliation{
    \institution{Hong Kong University of Science and Technology}
    \city{Oulu}
    \country{Hong Kong SAR}
}

\author{Simo Hosio}
\orcid{0000-0002-9609-0965}
\email{simo.hosio@oulu.fi}
\affiliation{
    \institution{University of Oulu}
    \city{Oulu}
    \country{Finland}
}

\renewcommand{\shortauthors}{Paananen, et al.}

\begin{abstract}
Recent advances in extended reality (XR) technologies have enabled new and increasingly realistic empathy tools and experiences. 
In XR, all interactions take place in different spatial contexts, all with different features, affordances, and constraints. We present a systematic literature survey of recent work on empathy in XR. As a result, we contribute a research roadmap with three future opportunities in XR-enabled empathy research across both physical and virtual spaces.


\end{abstract}

\ccsdesc[500]{Human-centered computing~Mixed / augmented reality}
\ccsdesc[300]{Human-centered computing~Graphical user interfaces}

\keywords{Extended Reality (XR), empathy, spatiality, the metaverse, Human-Computer Interaction (HCI)}

\maketitle

\section{Introduction}

Empathy can be characterized as an emotional response caused or related to another's emotional state~\cite{eisenberg1990empathy}. It serves society through understanding the situation of others, allowing humans to form more meaningful bonds and to act against injustices~\cite{elizabetha.segalAssessingEmpathy2017}. With the significant societal changes brought by the rise of computer systems, the empathic effect of digital media has been increasingly investigated. Accordingly, significant research has been conducted on eliciting empathy through virtual simulations. By leveraging immersive technologies such as extended reality (XR), these simulations enable the users to understand the experience of others regardless of their own temporal, cultural, demographic, or spatial context.


The way we experience empathy is highly related to our local context.
In the domain of ubiquitous computing and Human-Computer Interaction (HCI), context is generally understood as anything and everything that can characterize the situation of an entity. Context serves as the frame of reference through which we make sense of our world. Therefore, understanding context is essential for developing and analyzing computational systems~\cite{dourishWhatWeTalk2004}. Among the many components of context, space plays a central role~\cite{foucaultOtherSpaces1986a} by affecting our perception of the world and our experiences. Humans form meanings with spaces, such as safety and preference to some places over others~\cite{paananenInvestigatingHumanScale2021}. As such, being in a space leads to making and engaging with different contexts.
Spaces affect one's understanding of the situation, and thus how they experience empathy~\cite{elizabetha.segalAssessingEmpathy2017}. For instance, a person falling in the street has a different meaning to a person falling on a theatre stage.  Due to this human connection to spaces, empathy is necessarily bound to the spatial context in which they operate. 
Despite the spatial nature of humans and its significant relationship to context and empathy, space is still an understudied topic in HCI research~\cite{kirshArchitectsDesignersThink2019,kroghSensitizingConceptsSociospatial2017}. To help make sense of this rapidly emerging field of research, we present a systematic literature review on empathic XR applications under the consideration of their spatial contexts.

\subsection{Empathy}
Empathy is generally understood as the prosocial behavior that allows humans to feel and understand the experiences of others~\cite{elizabetha.segalAssessingEmpathy2017}. 
While the concept of empathy seems straightforward at surface level, 
its history and development are multifaceted. The concept originates from the notion of Einfühlung -- translating to ''feeling into" -- developed by German philosopher Robert Vischer, and later brought into psychology by Theodor Lipps in 1903~\cite{elizabetha.segalAssessingEmpathy2017}. In contrast to the interpersonal character trait as it is known today, Einfühlung is based on aesthetic theory to refer to the appreciation of objects and the relationship one has with their surroundings. 
In more recent literature, empathy carries a complex notion of behaviors and experiences revolving around undergoing or understanding another's experience.
Empathy thus encompasses nuanced concepts such as perspective-taking, compassion, sympathy, and pity.
Contradictory viewpoints exist as well: it has been suggested empathy is often simplified and misunderstood~\cite{bollmer_empathy_2017}. By proposing the concept of ''radical compassion" as an alternative to empathy, \citeauthor{bollmer_empathy_2017} argues for a more careful stance in assuming the experiences of the \textit{Other}. 


Empathy comprises two processes: affective empathy, related to feeling, and cognitive empathy, related to understanding~\cite{elizabetha.segalAssessingEmpathy2017}. Affective empathy has been shown to be more constant as it is related to the subject's personality, while cognitive empathy is easier to influence through external stimuli~\cite{pratteEvokingEmpathyFramework2021}. The recent notion of somatic empathy describes the embodied experience of empathy, for instance, wincing at another's experience of pain~\cite{raineCognitiveAffectiveSomatic2018}, and understanding the children's perspective of interacting with everyday objects that are primarily designed for adults~\cite{graspEmpathy}. This process has been a prevalent topic in cognitive neuroscience by investigating how the mirror neurons in human brains are a key element in the empathic process~\cite{elizabetha.segalAssessingEmpathy2017}.  


Empathy is heavily related to the person's own physical and social attributes, culture, and experiences. As such, it is difficult to measure and compare empathy between people. Factors such as race, gender, sexuality, social status, and culture affect how empathy is experienced. Generally, it is easier to form empathy between two people who share some aspects of their character \cite{pratteEvokingEmpathyFramework2021}. 
Over the years, multiple measurement instruments have been developed to assess empathic traits in people. The most popular self-report measures are the Interpersonal Reactivity Index (IRI), the Empathy Quotient (EQ), and the Questionnaire of Cognitive and Affective Empathy (QCAE) \cite{elizabetha.segalAssessingEmpathy2017}. Further, empathy can also be investigated on the neurological level. For instance, modern brain imaging technologies such as fMRI have been able to pinpoint the neurological basis of empathy in the mirror neuron system \cite{elizabetha.segalAssessingEmpathy2017}. However, neuroimaging technologies require specific expertise which is not often available to researchers outside of the field.

Empathy is thus a complex process that has behavioral, neurological, and cultural bases. As such, a robust and all-encompassing definition of empathy is difficult to formulate. 
The present paper considers ``empathy'' as a general term describing the process through which one person is ``feeling with'' another's experience.
Further, in the context of computer-mediated empathy described in this paper, a distinction should be made between empathic design and designing empathy tools.
Empathic design refers to understanding the user's experience of a product or a service during the design process. It is a typical approach for understanding the user in the domains of design and research~\cite{wrightEmpathyExperienceHCI2008}. 
On the other hand, empathy tools are artifacts designed to elicit an empathic response in users~\cite{pratteEvokingEmpathyFramework2021}. 
With the rise of extended reality technologies and their easier development processes, 
virtual empathy tools have become a popular way to investigate, elicit, and assess empathy~\cite{pratteEvokingEmpathyFramework2021}. This paper aims to survey the most prominent works on virtual empathy tools, as well as the observations into methods employed to measure empathy.
In testing the effectiveness of empathy tools, laboratory-based studies often take the advantage of simulations to elicit empathy in virtual situations. Conducting empathy studies in the wild brings difficulties in replicability. However, the effect of the designer in a laboratory experiment is a point of contention~\cite{pratteEvokingEmpathyFramework2021}. To form more robust practices, there is growing literature on different frameworks for empathy in technological contexts. For example, \citet{careyMeasuringEmpathyVirtual2017} proposed a framework for gathering data on empathy in VR studies, and the framework by \citet{w.b.owaisQuantifyingEmpathyVirtual2020} focused on quantifying empathy experiences in immersive virtual environments. We thus classify the studies based on the research methodology, to address the diverse measures.

\subsection{Empathy in Extended Reality}

\begin{figure}
    \centering
    \includegraphics[width=0.9\textwidth]{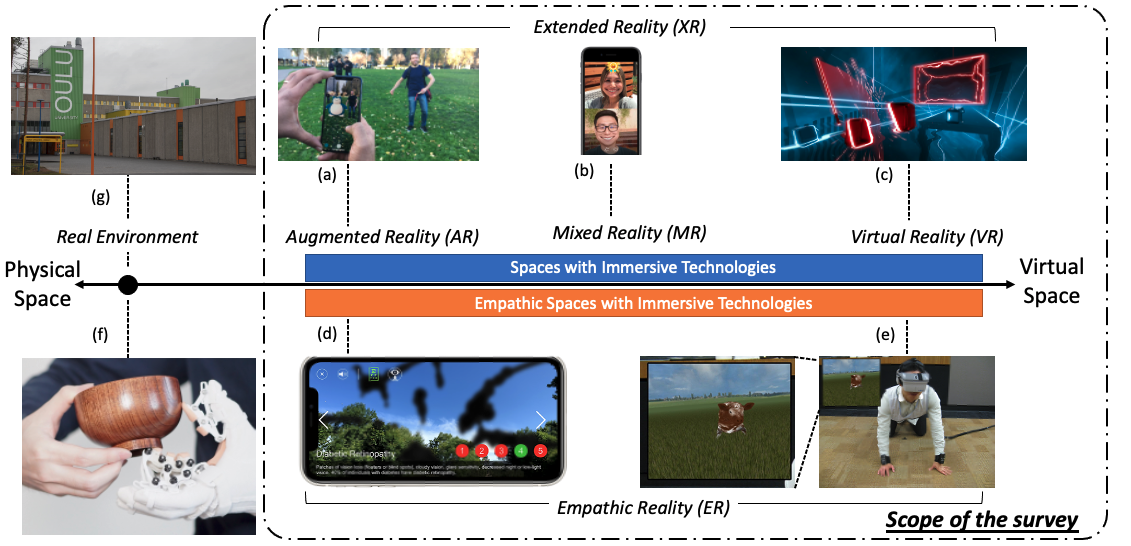}
    \caption{A revised spectrum of the virtual-physical continuum of virtual environments (Adapted from \citet{milgramTaxonomyMixedReality1994a}) known as Extended Reality (XR) -- (a) Pokemon Go; (b) Facebook Messenger; (c) Saber Fight; Furthermore, we added a layer of empathy on top of XR, with examples: (d) simulating visually impaired person with a hand-held AR display; (e) Body Transfer (Virtual Embodiment) as a cow~\cite{Ahn2016ExperiencingNE}; (f) A tangible wearable for emulating kids' grasping experiences to everyday objects, namely HandMorph~\cite{graspEmpathy}; (g) Real-world Environments. }
    \label{fig:VC_continuum}
\end{figure}

Virtual environments refer to computer-mediated 3D worlds characterized by immersion, realism, malleability, and scalability~\cite{csur-metaverse-2013}. In such environments, virtual entities can be either confined to purely virtual worlds (Virtual Reality, VR) or integrated within physical environments (Augmented Reality, AR) according to the well-recognized virtual-physical continuum by \citet{milgramTaxonomyMixedReality1994a} (the upper part in Figure~\ref{fig:VC_continuum}). In the middle of the continuum, the coexistence of physical objects and virtual entities construct virtual environments, known as mixed reality~\cite{Speicher-MR-CHI} (MR). The aforementioned virtual environments, regardless of the degree of virtuality and physicality, can be described by a collective term -- Extended Reality (XR: AR/MR/VR). Further, the immersive tools can also be used to remove or obstruct parts of perception, often visual, described as ``diminished reality'' (DR) \cite{moriSurveyDiminishedReality2017}.

Virtual environments can be regarded as facilitators of empathy, as depicted in a popular talk by \textit{Chris Milk} through advocating the prospect of virtual environments as \textit{the ultimate empathy machine}~\cite{milk_2015}. VR allows users to feel and understand the difficulties or struggles of other individuals~\cite{adefilaMyShoesFutureExperiential2016}, while AR can augment their surroundings with information of others~\cite{allowayGottaCatchEm2021}. Further examples under the virtual-physical continuum are available in Appendix~\ref{appendix:included_publications}. 

Alongside the development of virtual environments, has been the emergence of the metaverse~\cite{Lee2021AllON}, which describes diversified virtual environments in everlasting, gigantic, and wide-reaching ways. Although the metaverse is at the nascent stage~\cite{Lee2021AllON}, the prospect of the metaverse describes that the virtual entities will integrate into many aspects of our daily routines, and people will spend significant proportions of time with such virtual entities. Supported by several ambitious plans of metaverse cities, such as Seoul\footnote{\url{https://www.washingtonpost.com/world/asia_pacific/metaverse-seoul-virtual/2021/11/27/03928120-4248-11ec-9404-50a28a88b9cd_story.html}} and Shanghai\footnote{\url{https://www.cryptoglobe.com/latest/2022/01/shanghai-seeking-to-integrate-metaverse-over-next-five-years/}}, virtual environments can become indispensable in our urban spaces. With such prospects and anticipated user numbers, virtual-physical blended urban spaces will become a frontier of metaverse research. 

More importantly, such virtual-physical urban spaces should consider the diverse needs of stakeholders, meanwhile, urban designers can leverage virtual environments to inform concerns and resolve conflicts among stakeholders with conflicting needs in shared ways~\cite{Lee2022TowardsAR}. 
When virtual environments meet empathic design, we can foresee that empathy can play a vital role in multitudinous virtual environments. As shown in Figure~\ref{fig:VC_continuum} (the lower part), all virtual environments under the continuum of XR can be augmented by a new layer of empathy, namely \textit{Empathic Reality} (ER). Through the immersive (virtual) lens on top of our physical spaces, XR technologies promote alternative experiences to citizens along a pathway of empathic spaces. As so, 
as an extension of our physical reality, XR serves to bring new experiences on top of, or in place of our usual experiences\footnote{Note ubiquity of spatial metaphors in languages.}. This survey paper focuses on the whole range of XR technologies and their potential of connecting empathic elements and our physical spaces. 

\subsection{Humans in Spaces: Spatiality}
Context is tied with human experiences. In the process of understanding empathy, we 
analyze context through the lens of spatiality. \textit{Being and Time} by \citeauthor{heideggerBeingTime1985} is one of the foundational works on  spatial existence. In this book \citeauthor{heideggerBeingTime1985} claims human existence is first and foremost understood through the process of taking and making space. The fundamental nature of space for human existence can be seen in the spatiality of human language, memory, and experiences. Following a similar path,~\citet{bollnow_human_2020} elucidates the nuanced ways in which humans understand, feel, and embody spaces. On a personal level, meanings we attach to spaces can be positive or negative, influencing whether we seek or avoid such places. These places can be public or private or limited to certain social groups, e.g. residents, employees, or students. Spatiality describes the various ways we inhabit spaces, taking into account the full extent of spatial experience. Spatiality serves as a useful context through which we understand different phenomena, e.g., in the context of the current paper, where, and in what kind of XR spaces do we experience empathy?

\citeauthor{bollnow_human_2020} argues that the psychological mood and the atmosphere of the space are the same. A person's emotional state affects the perception of a space, which in turn affects the atmosphere the space elicits. The connection between the atmosphere of the place and empathy has also been made by others, e.g., through the metaphor of sphere of mutual affection by~\citet{sloterdijk_spheres_2016}, or as applied to geographical software by~\citet{widmer_foams_2020}. Subjects sharing a space are also interacting with each other. 
A study by \citet{maglioSpatialOrientationShrinks2014} found that people felt more similarity with people that they were spatially oriented towards \cite{maglioSpatialOrientationShrinks2014}. And so, as the psychological distance is not always equal to the geometric distance, we need to consider the experience of space as a co-created meaning. In other words, being in a space is a process of empathic interaction, where the space and its inhabitants are in a shared creation of mood. As this process is holistic and formed through the co-presence of people with their unique experiences of space, the atmosphere of place resists a singular definitive description.

In computing, spatial experiences were first brought into discussion by \citet{harrison_re-place-ing_1996}, by examining the notions of place and space in the context of computer-supported collaborative work (CSCW). 10 years later, \citet{dourishRespaceingPlacePlace2006} returns to the topic to address conceptual difficulties the spatial notions carry in the face of developing technology. Building on the works of \citet{harrison_re-place-ing_1996}, \citet{lentini_space_2010} develop five dimensions for understanding the experience of space: geographical, sensorial, cultural, personal, and relational. Through these dimensions, the richness of experiences in the urban environment is more readily understandable for developing ubiquitous computing systems. Similarly, proxemics, the study of interpersonal distances, was brought to HCI through the analysis of social and spatial contexts of technology \cite{kroghSensitizingConceptsSociospatial2017} to highlight how modern technologies lack in terms of spatiality. Following the rising popularity of spatial focus, the recently developed notion of Human-Building Interaction (HBI) aims to understand what challenges and opportunities are in the cross-section of architecture and HCI \cite{alaviIntroductionHumanBuildingInteraction2019}. As a further elucidation of HCI and architecture, \citet{kirshArchitectsDesignersThink2019} describes how the fields differ in understanding what interaction means, and HCI could serve well to employ more networked modes of interaction. At the core of these modes of understanding spaces is the social quality of spaces \cite{kirshArchitectsDesignersThink2019}. To this end, understanding spatial experiences from the perspective of everyday spaces has been brought up as a relevant research focus in HCI~\cite{paananenInvestigatingHumanScale2021}.
Immersive technologies such as augmented and virtual reality open new opportunities to study spatiality in HCI. The immersive capabilities of VR enable users to experience different spaces, while AR augments everyday spaces with digital content. The recent focus on the metaverse serves as a way to produce persistent spaces in the virtual world, mirroring the relative stability of the physical environments.


\subsection{Previous Surveys}
XR, empathy, and spatiality have all been subjected to various literature surveys in the past. However, the specific intersection of these which is pivotal to empathy research has not been examined earlier. In a recent survey, \citet{pratteEvokingEmpathyFramework2021} surveyed the use of empathy tools in HCI research. From a set of 26 publications, they clarified the language around the audience of empathy tools, analyzed used technologies, and the role of agency, perspective, and sensation for fostering empathy. Finally, they presented recommendations for HCI research for empathy tools. To build on this work, in the present survey we focus on the spatial context of different empathy tools. In the context of caregivers, \citet{brydonVirtualRealityTool2021} reviewed 7 studies using VR empathy tools. The findings supported using VR technologies to elicit empathic behavior in caregivers. In our approach, we also consider XR technologies. Using VR nonfiction (VRNF) media, \citet{bevanCurtainUltimateEmpathy2019} analyzed 150 VNRF titles to highlight what narrative techniques were used to elicit empathy. The findings suggest the relatively new media is experimenting with suitable paradigms for interaction, viewer roles, and more. In contrast, we target XR applications that can also have a non-narrative focus. In surveying VR empathy tools, \citet{w.b.owaisQuantifyingEmpathyVirtual2020} formulated a framework for immersive virtual reality environments. However, the scope was quite limited and is not representative of the current state of research. Moving into the spatial realm, \citet{danyluk_design_2021} reviewed how virtual reality tools were used to present and interact with Worlds-in-Miniature (WiMs), which are interactive representations of things ranging from small objects to a planetary scale. The multiple scales afforded by the XR technologies provide a shift in a spatial context, allowing new perspectives. These spatial contexts serve useful lens for the empathic focus in our survey. \citet{Lee2022TowardsAR} surveyed the urban spaces as the gigantic immersive scenarios with augmented reality, and further brought up the consideration of various stakeholders in such immersive urban spaces. However, their works neglect the empathy in such physical spaces that is evoked with immersive technologies. Finally, \citet{Paiva2017} surveyed research on virtual agents and robots with regards to computational empathy. They highlight how virtual agents can serve different roles in empathic interaction, which in turn affects how they are perceived. In contrast, we are not as focused on computational empathy, but rather the tools that can be used to elicit empathy.

\subsection{Scope of the Present Survey}
This survey reviews recent work on how empathy, spatiality, and XR technologies together enable novel forms of empathy research and experiences. We present the state of current empathy research that uses XR technologies and how different spatial contexts are used in these approaches. As the outcome of the survey, we contribute a research roadmap that builds on identified opportunities within the intersection of the reviewed fields.

We answer the following research questions in the context of empathy research using XR technologies:

\begin{enumerate}
    \item What tools or technologies are used and how?
    \item Who is being empathized and by whom?
    \item What perspectives are taken in empathy?
    \item In what spatial contexts is empathy elicited?
\end{enumerate}

Section~\ref{section:methodology} describes our systematic review process and data extraction. In Section~\ref{section:synthesis}, we describe the different themes brought up through the synthesis of data extraction. Section~\ref{section:discussion} describes the research opportunities 
we identified and discusses an array of related issues. In Section~\ref{section:research_roadmap}, we outline a research roadmap for spatial empathy tools that leverage XR technologies. Finally, Section~\ref{section:conclusion} concludes the survey.

\section{Review Methodology} \label{section:methodology}
To conduct the literature survey systematically. we applied the PRISMA methodology as described by \citet{page_prisma_2021}. Web-based tool \textit{Covidence} was used to handle the process in collaboration with multiple authors.

\subsection{Search Strategy}
\subsubsection{Keywords}
The chosen keywords were in the three dimensions of empathy, spatiality, and XR technologies. To take into account the different scales and contexts of spatiality, keywords "urban", "space", "place", "building", "city", and "environment" were used. Empathy was focused on with the related terms "empathy", "sympathy", "pity", and "compassion". Finally, to cover the wide range of different XR technologies we targeted "VR", "AR", "MR", and "DR" -keywords. The keywords and the resulting general search string is described in Table~\ref{table:search_keywords}.

\begin{table}
    \centering
    \small
    \caption{Chosen keywords and the resulting search string for the database searches.}
    \begin{tabular}{l p{10cm}}
        \hline
        XR keywords      & (virtual / mixed / augmented / diminished) reality       \\
        Empathy keywords & empathy / sympathy / pity / compassion                   \\
        Spatial keywords   & urban / space / place / building / city / environment    \\ 
        \hline
        Search String    & ("empathy" OR "sympathy" OR "pity" OR "compassion") AND ("urban" OR "space" OR "place" OR "building" OR "city" OR "environment") AND ("virtual" OR "mixed" OR "augmented" OR "diminished") AND "reality" \\ 
        \hline
    \end{tabular}
    \label{table:search_keywords}
\end{table}

\subsubsection{Databases}
We primarily used ACM Digital Library (ACM DL) and IEEE Xplore Digital Library to focus on technology- and computer science -related literature. For a more thorough review, we also included Web of Science, Scopus, EBSCOHost, and Elsevier (ScienceDirect) to cover fields such as architecture and psychology that relate to spatial contexts and empathy. Using the previously described keywords, we formulated the search strings for the specific databases, as described in the Appendix~\ref{appendix:search_strings}, alongside the search results. We note that Elsevier allowed only 8 Boolean operators in the search string, reducing the query length. Also, for ACM DL we targeted the search string to the Title, Abstract, or Keywords for more relevant results.

\subsection{Data Extraction}
To extract relevant information from the included articles, we developed a data extraction rubric. Initially, the first author of this survey selected 10 articles pseudo-randomly (sorted by author name: \cite{calvertImpactImmersingUniversity2020, brownVirtualRealityClinicalexperimental2020, baiEmpathyToolDesignEye2019, salminenBioadaptiveSocialVR2018, allowayGottaCatchEm2021, adefilaMyShoesFutureExperiential2016, kromaAlzheimerEyesChallenge2018, korsBreathtakingJourneyDesign2016, bouchardEmpathyVirtualHumans2013, bujicEmpathyMachineHow2020a}), and after reading the articles, developed items around the relevant aspects of XR technologies, empathy, and spaces. The initial data extraction rubric was then evaluated by all authors and finalized into the rubric described in the Table~\ref{tab:DE_criteria}. Using the finalized rubric, the first and second authors conducted the data extraction for each article, and for any conflicts, consensus was formed through discussion. 

The data extraction items DE1 -- DE4 are used for general descriptors for the papers -- study ID, title, publication type, and keywords. The XR technologies (DE5) described the most general types: AR, VR, VR video (non-interactive), MR, CAVE, and other. Interaction (DE6) modalities show the range of used hardware and how they can be used in the virtual environment. Locomotion (DE7) describes how the user moves in the virtual environment, whether through their actions (e.g., using a controller), or through a scripted movement (e.g., moving video). Measuring empathy (DE8) was focused on the general scientific measurement methods of qualitative, quantitative, and mixed, and also N/A in the case empathy was not measured or unclear. Measuring the longer-term effects of empathy (DE9) has been recognized as a good practice for understanding the impact of empathy tools \cite{pratteEvokingEmpathyFramework2021}. To clarify the language around the subject and object of empathy, we use ``user'' for the person using the empathy tool and ``\textit{Other}'' for the human, animal, or entity we are empathizing. As such, embodiment for the \textit{Other} (DE10) and the user (DE11) were categorized by general types of human, animal, robot, object, and other. Concepts such as ``nature'' or ``sustainability'' were marked as ``other''. The items DE12 and DE13 describe how the empathy experience is presented to the user. The point of view (POV) compared to the user (DE12) refers to how objectively the experience is presented, or how much control the user has for experiencing the story (subjective). The narrative point of view (POV) of the user (DE13) categorizes how wide understanding of the situation the user is given. The items DE14 and DE16 are related to the context of the situation, through the notion of proxemics, the study of distances \cite{hallHiddenDimensionMan1969}. The type spaces in the experiences (DE14) describe the qualities of the different proxemics distances. To exemplify these, being a remote observer is public, conversation with someone is social, focusing on your near surroundings is personal space, and the intimate is reserved to the spaces most closest to you. Distance in the empathic interaction (DE16) is based on how far the user is from the \textit{Other}. Empathizing perceptual experiences (i.e. vision or hearing impairment), were marked as ``intimate''. Another dimension of the quality of spaces is captured by the privacy of the spaces (DE15), ranging from public to private. A semi-public space is a space that is meant for the public, but the access is partially limited (e.g. a shopping mall), and semi-private space is limited to its owner but is occasionally accessed by others by permission (e.g. a backyard). The \textit{Other} of empathic interaction (DE17) refers to who, and with how many people, are experiencing the empathy. One-to-one means the user is experiencing a single person's experiences in the virtual environment. Accordingly, many-to-many is used for situations where several people are participating in the experiment, and are empathizing many Others.

\begin{table}
    \caption{Data extraction rubric for the selected papers.}
    \small
    \begin{tabularx}{\textwidth}{l p{0.42\textwidth} p{0.42\textwidth}}
        \toprule
        ID & Data Extraction & Type \\ \midrule
        DE1         & Study ID & Open text \\
        DE2         & Title & Open text \\
        DE3         & Publication type & Conference paper, journal article, thesis, other \\
        DE4         & Keywords & Open text \\
        DE5         & Used XR technologies & AR, VR, VR video (non-interactive), MR, CAVE-system, other \\
        DE6         & Interaction type & Gesture, touch, keyboard/controller/joypad, other \\
        DE7         & Locomotion & Yes, no \\
        DE8         & How was empathy measured? & Quantitatively, qualitatively, mixed methods, N/A \\
        DE9         & Did they measure longer term effects of empathy? & Yes, no, other \\
        DE10        & Who/what is the \textit{Other}? & Human, animal, robot, object, other \\
        DE11        & What is the user embodying in the empathic process? & Human, animal, robot, object, other \\ 
        DE12        & What is the point of view of the experience compared to the user? & Objective, subjective, omniscient \\ 
        DE13        & What point of view is used? & First POV, fly-on-the-wall, omniscient, other \\ 
        DE14        & What type of spaces the empathy exists in? & Intimate, personal, social, public space, other \\ 
        DE15        & How public/private is the space? & Public, semi-public, semi-private, private, other \\ 
        DE16        & Distance in the empathic interaction? & Intimate, personal, social, public distance \\ 
        DE17        & The \textit{Other} of empathic interaction & One-to-one, one-to-many, many-to-one, many-to-many, other \\ \bottomrule
    \end{tabularx}
    \label{tab:DE_criteria}
\end{table}

\subsection{Survey Results}
\begin{figure}
    \centering
    \includegraphics[width=0.7\textwidth]{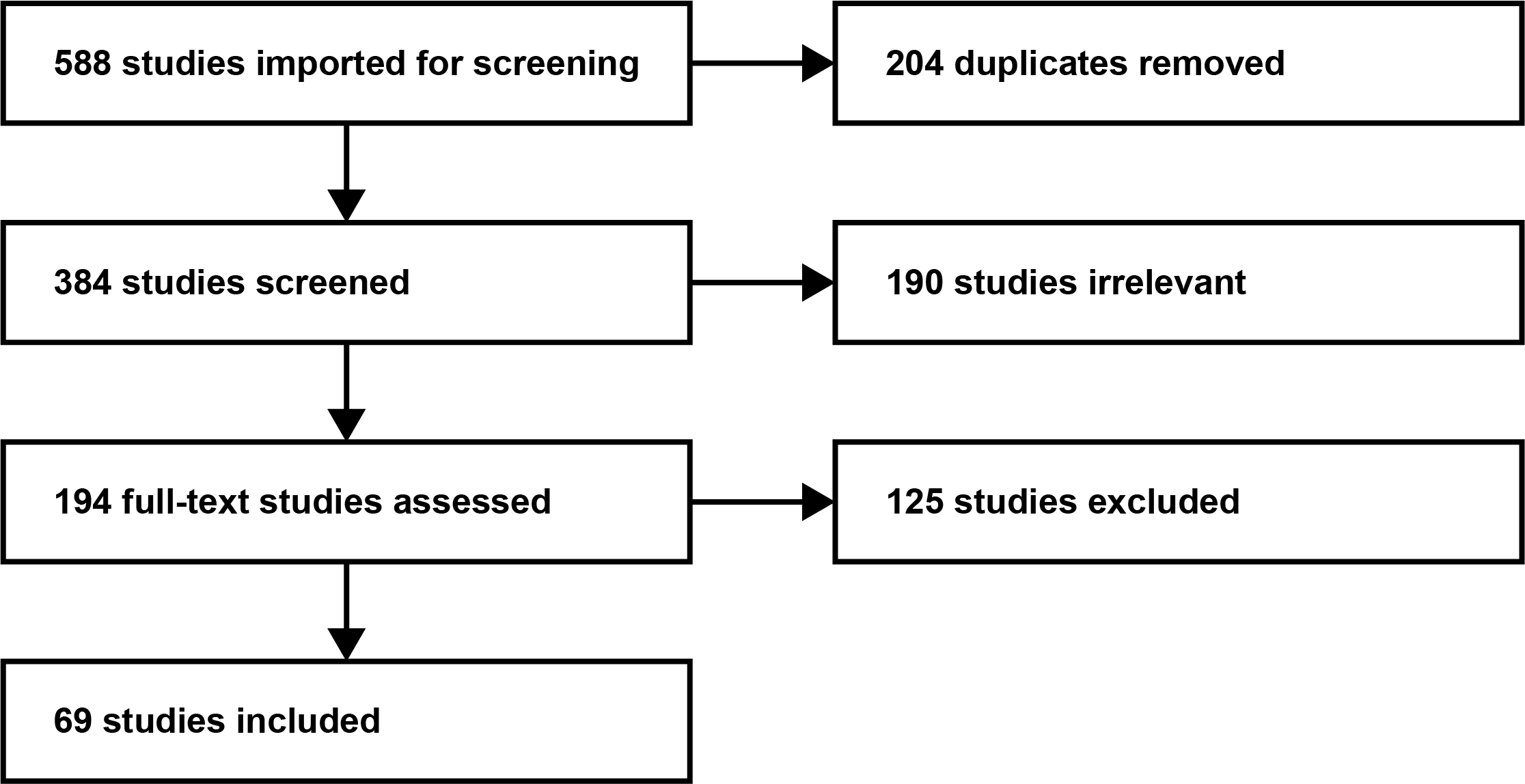}
    \caption{Systematic review process using PRISMA.}
    \label{fig:PRISMA}
\end{figure}

The outline of our systematic review process using PRISMA is shown in Figure~\ref{fig:PRISMA}. After completing the search, we started the screening process with 588 articles, of which 204 articles were removed as duplicates. We conducted the title and abstract screening process for 384 articles, of which 190 were excluded. After the full-text assessment, we excluded further 125 articles. The justifications and the counts (in parenthesis) for excluding the 125 papers are as follows: no XR technologies (17), is a survey paper (15), no spatial context (14), not a paper (14), duplicate (13), a proposal (13), no focus on empathy (13), no empirical study (12), empathy only a side-effect (5), only a framework (5), full-text not found (2), is a book (1), and non-English (1). Finally, we were left with 69 papers on which we conducted the data extraction.

\subsection{Description of the Included Articles}
\begin{figure}
    \centering
    \includegraphics[width=0.8\textwidth]{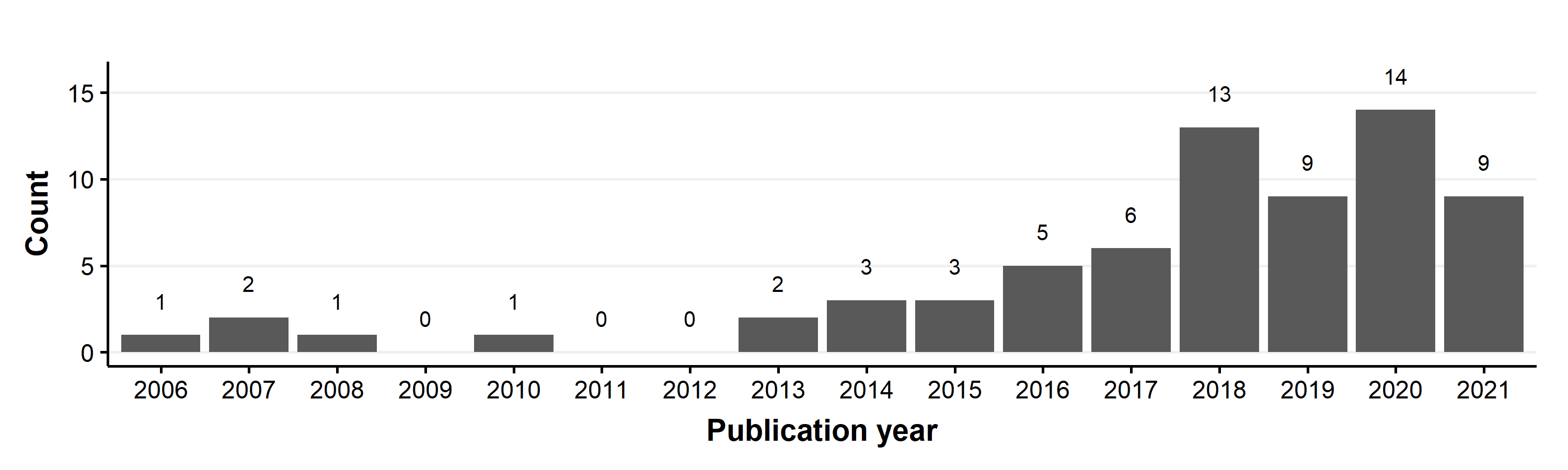}
    \caption{Articles per year (N=69)}
    \label{fig:articles_per_year}
\end{figure}

We extracted relevant information from the 69 articles that resulted from the screening process. This information was then analyzed and summarized statistically and graphically. Further qualitative insights were brought up in the iterative analysis process. The full list of the included articles is presented in Appendix~\ref{appendix:included_publications}. The articles were from years between 2006 and 2021, and the number of publications per year is shown in Figure~\ref{fig:articles_per_year}. It seems there is a general increase in the focus of our survey, however, aspects such as an increase in XR technologies, digital archival of research, and global scientific output can confound the results. As for the publication type, we had 42 (61\%) journal papers, 26 (38\%) conference papers, and 1 (1\%) thesis. As our data extraction took place in mid-2021, the results for that year are not representative of the whole year.

\begin{figure}
    \centering
    \includegraphics[width=0.9\textwidth]{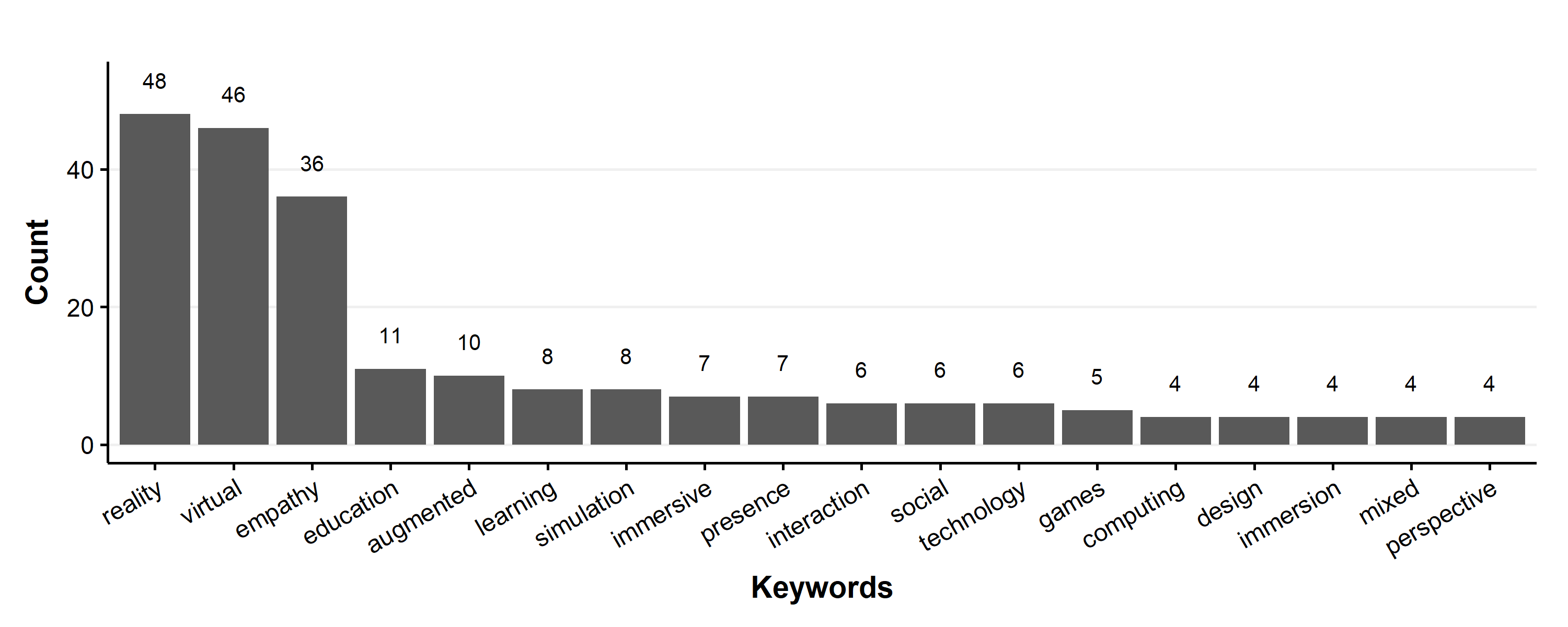}
    \caption{Keywords with frequency > 3}
    \label{fig:frequent_keywords}
\end{figure}

\begin{figure}
    \centering
    \includegraphics[width=0.6\textwidth, trim={0 6cm 0 6cm}, clip]{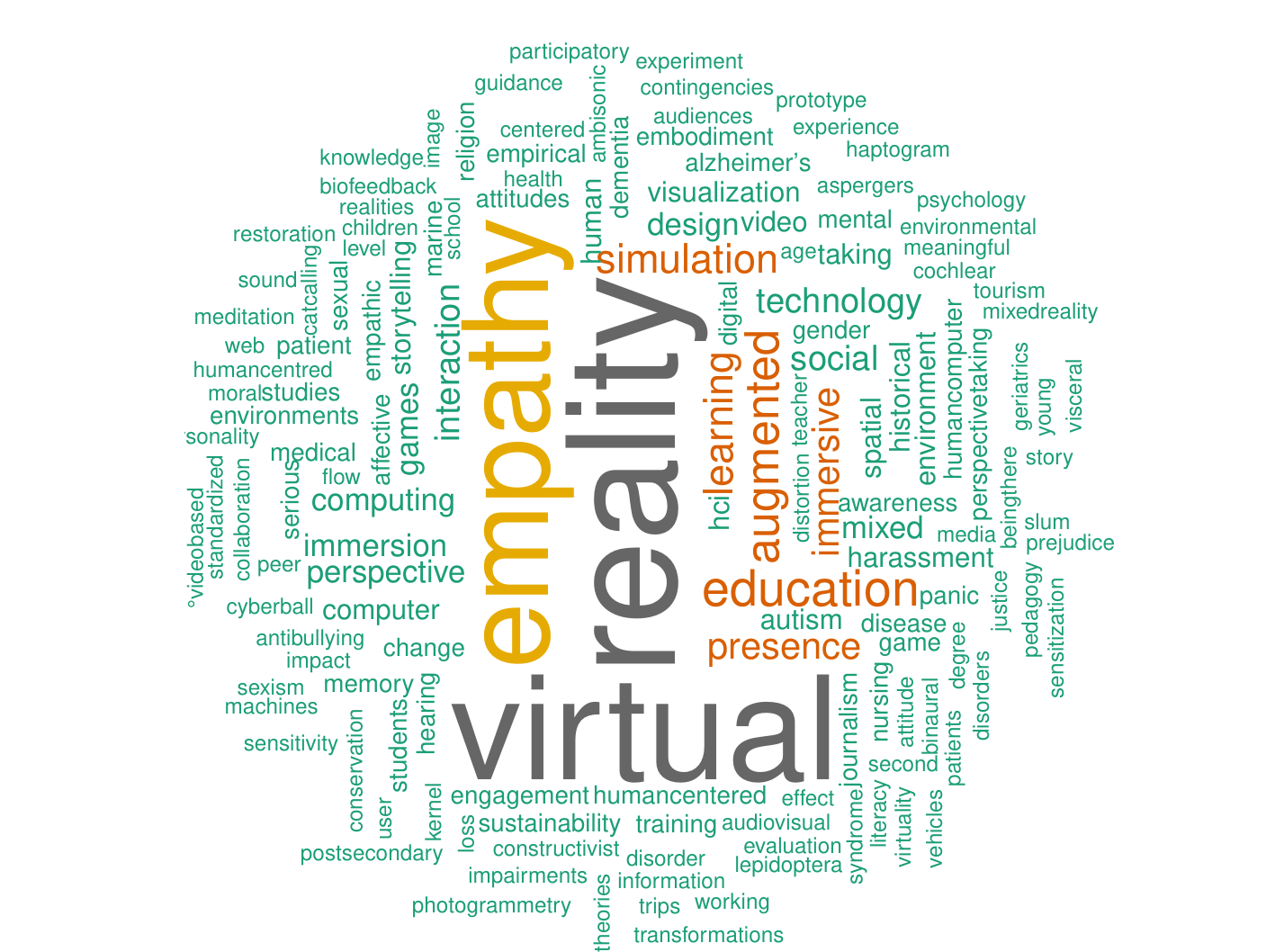}
    \caption{Used keywords in the articles}
    \label{fig:keyword_wordcloud}
\end{figure}

The articles come from various scientific disciplines: besides computer science, fields of medical, educational, and journalism science had employed XR devices for empathy. Most frequent keywords mention "reality", "virtual", and "empathy", after which the frequency of the keywords falls significantly, as shown in Figure~\ref{fig:frequent_keywords}. The wide range of the keywords is also visualized in Figure~\ref{fig:keyword_wordcloud}. While the keywords represent the themes of XR technologies and empathy, the theme of spatiality is not immediately present. This can be an effect of spatiality serving as the context for experience and is often less focused on, rather than the experience itself.

\section{Data Synthesis: XR Spaces for Empathy} \label{section:synthesis}
The included 69 papers cover a wide range of topics in empathy and XR research. Alongside the domain of HCI, empathy tools are applied fields such as medical \cite{deladismaMedicalStudentsRespond2007}, psychological \cite{christofiVirtualRealitySimulation2020}, educational \cite{calvertImpactImmersingUniversity2020}, journalism \cite{steinfeldBeThereWhen2020}, media \cite{gillathWhatCanVirtual2008}, religious \cite{johnsonUsingVirtualReality2018}, and geographical sciences \cite{griffinExploringVirtualReality}.
The following sections describe how our three research themes are present in the articles, leading into the discussion, where we describe the general trends and approaches for fostering empathy with XR.

\subsection{Technologies}
\begin{figure}
    \centering
    \includegraphics[width=\textwidth]{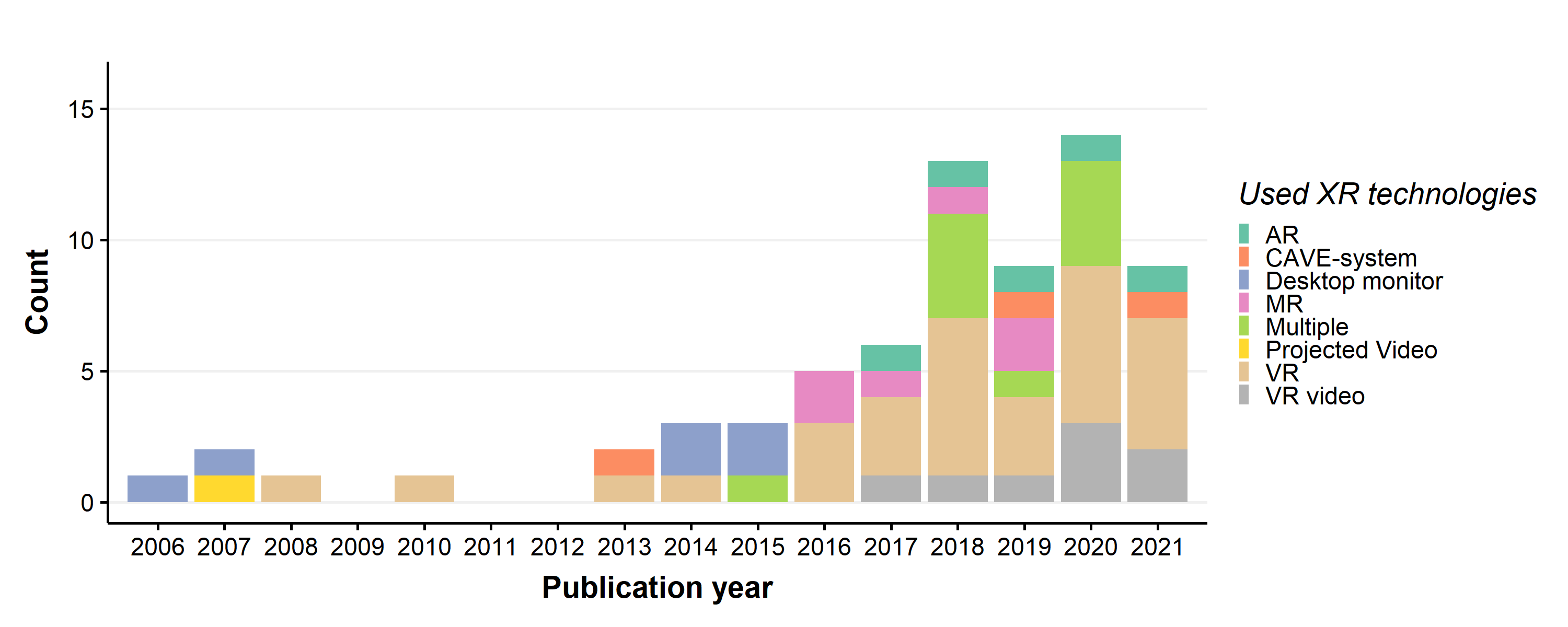}
    \caption{Used XR technologies by the year}
    \label{fig:used_technology_by_year}
\end{figure}

Figure~\ref{fig:used_technology_by_year} shows the usage of different XR technologies over the years. Using XR for fostering empathy has been increasingly popular, with VR being the most popular technology. Following the release of high-end headsets such as Hololens in 2016, AR technology has been steadily investigated, and studies using multiple technologies have seen a significant increase in popularity.
Specifically, VR is the most used technology (43\%), followed by VR video (12\%), desktop monitor (9\%), and AR (7\%) approaches. There is also a significant amount of papers (14\%) using multiple technologies, to assess the empathic effect across different media. For instance, several papers performed their studies' experiments on different platforms, such as VR and VR video (e.g., \cite{sweeneyUsingAugmentedReality2018a, christofiVirtualRealitySimulation2020, bujicEmpathyMachineHow2020a}).

The studies using multiple technologies can be categorized in two ways: studies using technologies with different levels of immersion, and studies where technologies with comparable complexity were used. As an example of technologies of different levels of immersion, \citet{bujicEmpathyMachineHow2020a} used a 360 HMD video, a 360 video viewed on a desktop monitor, and an internet article read on a monitor, i.e., high, medium, and low immersion, respectively. Similarly, \citet{p.r.jonesDegradedRealityUsing2018} developed and tested AR and VR technologies for eliciting empathy on visual impairments. As an example of non-comparative technology-induced empathy, \citet{stoverSimulatingCubanMissile2007} employed computer simulation and reading tasks to foster empathy. Furthermore, in the perspective-taking process of understanding the Cuban Missile Crisis, experimenters directed the study participants, students, to employ an active approach to write about their viewpoints.  

Mixed reality studies used consumer-market hardware such as Microsoft Hololens \cite{baiEmpathyToolDesignEye2019, piumsomboonEmpathicMixedReality2017}, or semi-custom HMD solutions with additional modalities \cite{werfelEmpathizingAudiovisualSense2016, leeWhoWantsBe2019, kromaAlzheimerEyesChallenge2018, korsBreathtakingJourneyDesign2016} to elicit empathy. For example, \citet{leeWhoWantsBe2019} built an ``unconventional driving machine'', where the user lies in a steel frame of a small buggy. A 3D camera and light detection and ranging (LiDAR) sensors provide visual feedback to the user through a VR HMD. 
As such, drivers or vehicle designers can understand the perspective of an autonomous vehicle and how it impacts the urban environment and other people. In a similar approach, \citet{korsBreathtakingJourneyDesign2016} designed an experience around the user's sensory system through audio-visual, touch, and olfactory stimuli. The experience of being a refugee hiding in a truck is enlivened by seating the user in a small wooden crate and accentuating the virtual visual feed with smells, touches, and sounds.

\begin{figure}
    \centering
    \includegraphics[width=\textwidth]{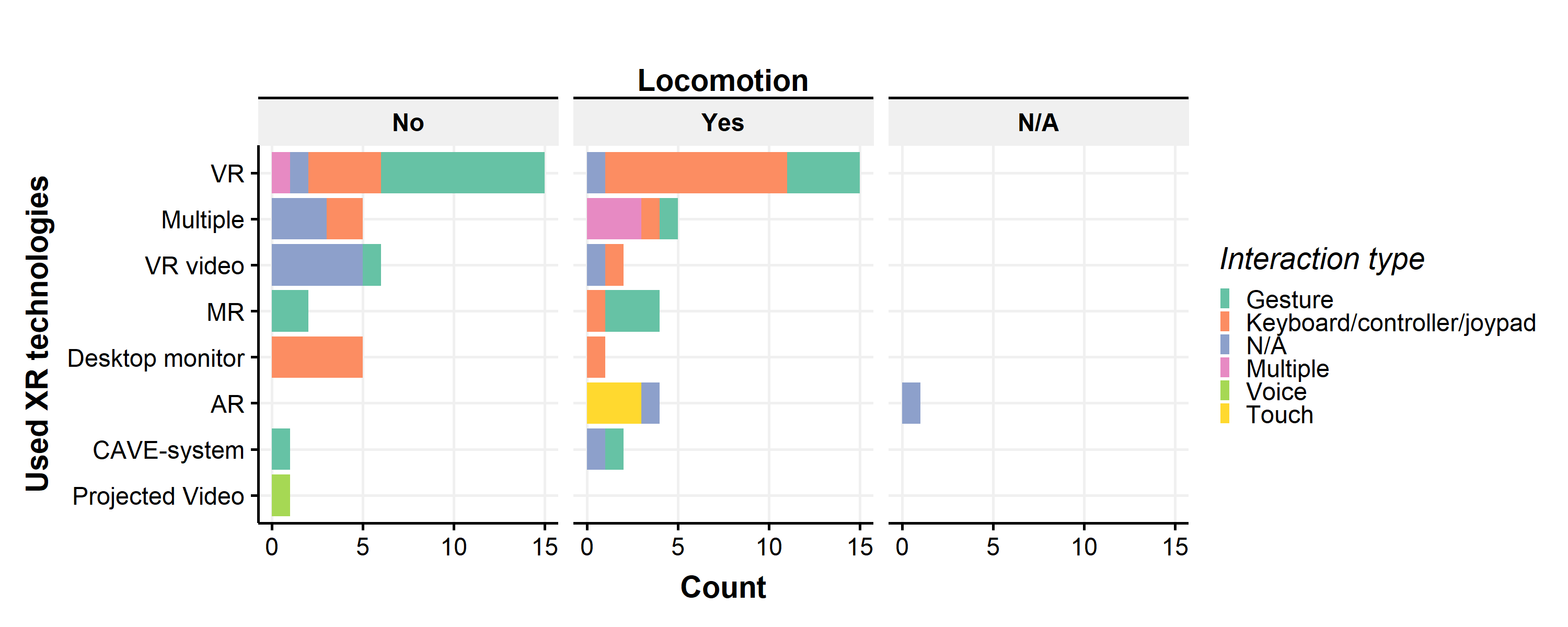}
    \caption{Used XR technologies by interaction methods, grouped by locomotion.}
    \label{fig:used_technology_by_interaction_type_by_locomotion}
\end{figure}

The overview of XR technologies, interaction paradigms, and locomotion in the studied works is shown in Figure~\ref{fig:used_technology_by_interaction_type_by_locomotion}. Interaction describes how the users were able to change the state of the world through their actions and what methods facilitated it. The chosen XR technologies partially dictate interaction methods, and VR is the most popular choice for eliciting empathy, using gestures and devices for interaction. In contrast, VR video describes experiences where the user is witnessing a filmed story, and as such, interaction is limited. However, \citet{raduUsing360VideoVirtual2021} used controllers to progress the story in a narrative VR video experience. As such, the interaction enables the user to progress the story at their own pace. Furthermore, we observe that many of the VR video experiments did not use locomotion, i.e., were focused on the user being a stand-still observer.

\subsection{Elicited Empathy}

83\% of the papers included in the study have used a form of a quantitative assessment of empathy, such as the self-administered IRI or QCAE -questionnaires, or a subset of them. Around 14\% of the papers focused on qualitative assessments of empathy, such as interviews \cite{embolHearMeVirtualRealityUsing, kellyOurDogProbably2020} or analysis of behaviour \cite{schaperDesignVirtualHeritage2017, kleinsmithUnderstandingEmpathyTraining2015}. One-third of the papers used both quantitative and qualitative methods to understand the empathic effect. Further, 17\% of the papers did not measure empathy and often focused on some forms of empathy tool prototypes \cite{claybornLivingRoomConservation2019}, or the topic was on empathy, but the measurement instruments did not include empathy \cite{haddodFeelAutismVR2019}.

\begin{figure}
    \includegraphics[width=0.8\textwidth]{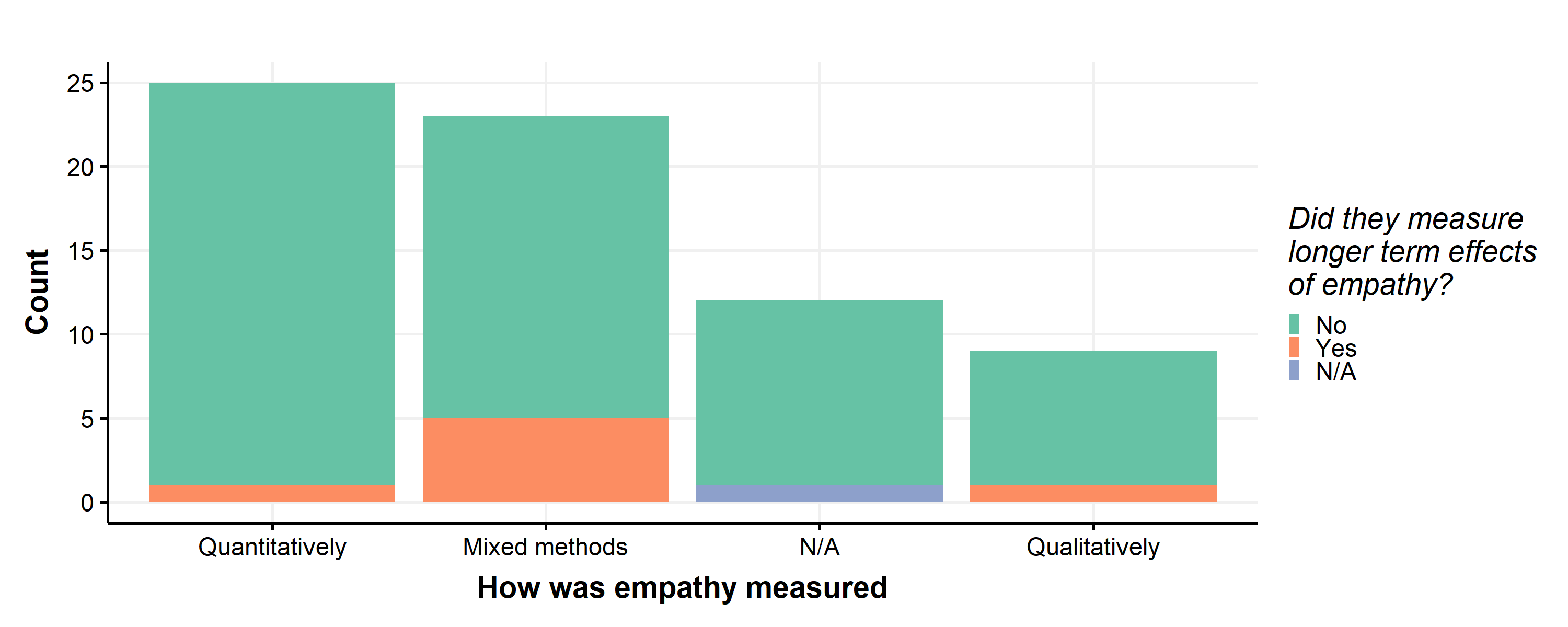}
    \centering
    \caption{Empathy measurement methods and longer-term measurements.}
    \label{fig:empathy_measurement_by_longer_term_measurement}
\end{figure}

10\% of the included studies measured the long(er)-term effects of empathy in their experiments. As shown in Figure~\ref{fig:empathy_measurement_by_longer_term_measurement}, longer-term effects were mostly measured in studies, which used mixed methods to measure empathy. For example, \citet{adefilaMyShoesFutureExperiential2016} interviewed a subset of the users after an empathy experiment on dementia, and \citet{herreraBuildingLongtermEmpathy2018} focused on the effect of empathy two, four, and eight weeks after the experiment. Out of the 7 papers measuring long-term empathy, 3 articles focused on educational contexts. For instance, \citet{efstathiouInquirybasedAugmentedReality2018} found the AR learning condition leads to greater historical empathy as compared to the non-technological approach. Furthermore, out of the 7 papers that measured long-term empathy, 4 articles reported a longer-term qualitative or quantitative difference in empathy from using the empathy tools (\cite{efstathiouInquirybasedAugmentedReality2018, tsaiInclusionThirdpersonPerspective2021, abadiaEffectivenessUsingImmersive2019, griffinExploringVirtualReality}). However, as the aforementioned studies did not use similar methods, it is not possible to compare the long-term effectiveness between the different approaches.

Most frequently, the users empathized with humans among all the studies we examined. In 84\% of the papers, empathy was directed towards humans. In contrast, 13\% of the papers focused on empathizing objects and 3\% animals. As an example of empathy towards animals, \citet{claybornLivingRoomConservation2019} designed a VR experience where the user completes tasks to understand the living relationships of butterflies, plants, and exotic species. Similarly, \citet{kellyOurDogProbably2020} built an AR tool to show the world through dogs' eyes, with their limited color accuracy. Objects or concepts that were emphatized included self-driving cars \cite{leeWhoWantsBe2019}, historical events \cite{sweeneyUsingAugmentedReality2018a, schaperDesignVirtualHeritage2017}, and sustainability practises \cite{munteanCommunicatingSustainableConsumption2020, l.calviVRGameTeach2017, m.poslusznyPromotingSustainabilityVirtual2020, mcmillanVirtualRealityAugmented2017}. 

\begin{figure}
    \includegraphics[width=0.8\textwidth]{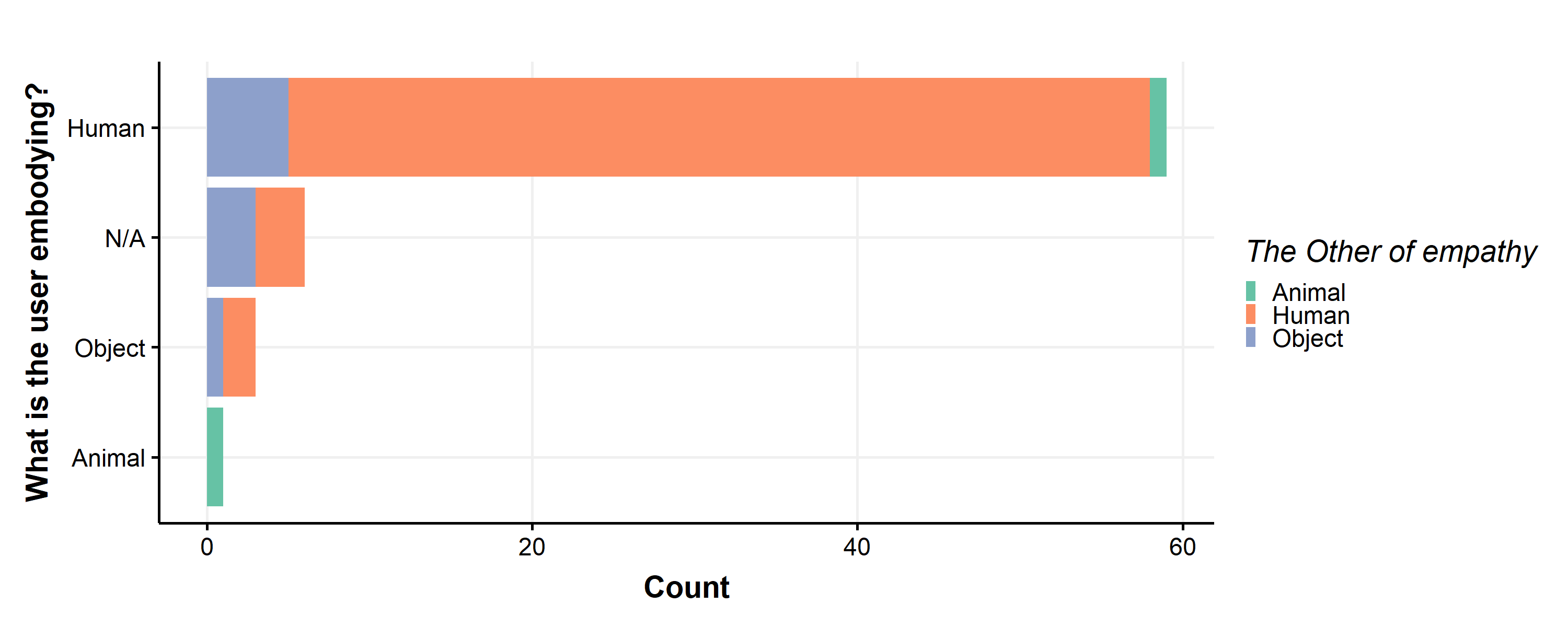}
    \centering
    \caption{What is the user embodying compared to the \textit{Other}}
    \label{fig:embodying_by_target}
\end{figure}

Embodiment follows a similar distribution as for the \textit{Other}: in 86\% of the papers, users embodied a human, in 4\% an object, and in 1\% of the studies they embodied in an animal. It is worth mentioning that 9\% of the papers had no clear embodiment and we mark them as N/A, e.g., in the case of many virtual 360 videos where a physical body is not visible. The relationship between the \textit{Other} and the embodiment is shown in Figure~\ref{fig:embodying_by_target}. In two papers, the users experienced from a non-human perspective to understand a non-human perspective: seeing the world through a dog's eyes \cite{kellyOurDogProbably2020}, and self-driving cars by acting as one \cite{leeWhoWantsBe2019}. While there were studies that used multiple human perspectives \cite{stavrouliaDesigningVirtualEnvironment2018}, no studies were using multiple types of embodiment.

The empathy experiments were mostly subjective (54\%) or objective (45\%). However, one study by \citet{shinEmpathyEmbodiedExperience2018} used an omniscient viewpoint, where the users viewed a 360 documentary describing the effects of war on children. As such, the stimulus presents multiple experiences of the children in the documentary. 


First POV is the most used viewpoint (91\%) to show experiences through the eyes of another. Four studies used the fly-on-the-wall viewpoint \cite{tettegahCanVirtualReality2006, bujicEmpathyMachineHow2020a, mcevoyEyesBystanderUnderstanding2015, e.mavromihelakiCyberball3D3DSerious2014}. A study by \citet{griffinExploringVirtualReality} used the omniscient POV to show the effects of slum tourism through a documentary. Both the first and third-person perspectives were used in the study by \citet{tsaiInclusionThirdpersonPerspective2021}, which aimed at improving emotion recognition in children with autism through a CAVE-like system.

\subsection{Spatial Context}
\begin{figure}
    \includegraphics[width=\textwidth]{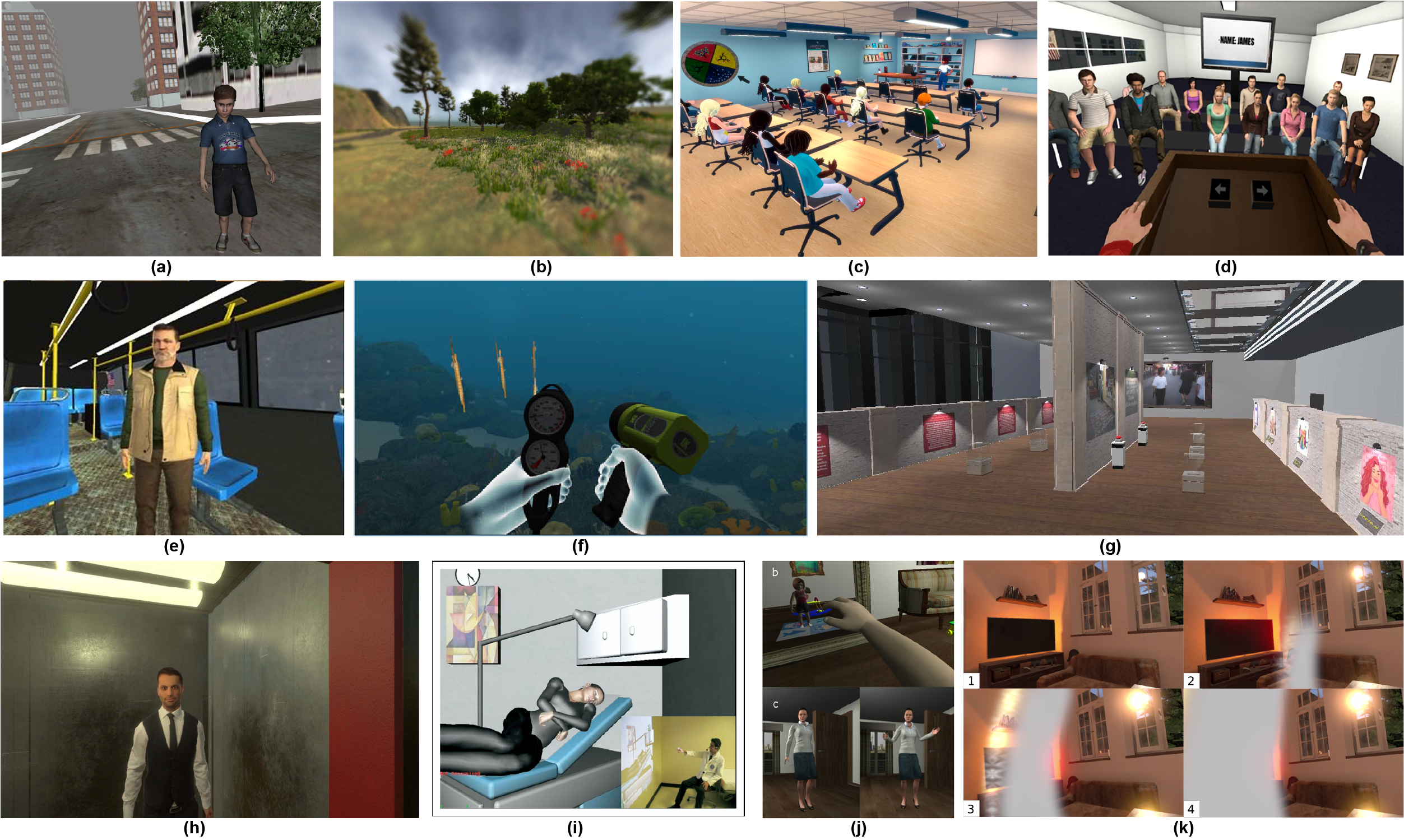}
    \centering
    \caption{Selection of various spatial contexts in the included papers, ranging from public (top left) to private (bottom right), as follows: 
    a) urban environment \cite{rosenbergVirtualSuperheroesUsing2013}, 
    b) nature \cite{g.m.rodriguezImmersiveApproachVisualizing2017}, 
    c) classroom \cite{embolHearMeVirtualRealityUsing}, 
    d) workplace \cite{vanloonVirtualRealityPerspectivetaking2018}, 
    e) public transport \cite{herreraBuildingLongtermEmpathy2018}, 
    f) underwater \cite{mcmillanVirtualRealityAugmented2017}, 
    g) museum \cite{helgertStopCatcallingVirtual2021}
    h) elevator \cite{v.russellHAVEExperienceInvestigation2018},
    i) doctor's office \cite{deladismaMedicalStudentsRespond2007},
    j) home \cite{hamilton-giachritsisReducingRiskImproving2018}, and
    k) home \cite{s.misztalSimulatingIllnessExperiencing2020}.}
    \label{fig:spaces}
\end{figure}

The included publications cover a large range of spatial contexts in which empathy is elicited. For instance, the studies used environments and spaces such as religious sites \cite{johnsonUsingVirtualReality2018}, slums \cite{griffinExploringVirtualReality}, museum \cite{nechitaAugmentingMuseumCommunication2019}, library \cite{raduUsing360VideoVirtual2021}, cafe \cite{felnhoferPhysicalSocialPresence2014a}, school \cite{mcevoyEyesBystanderUnderstanding2015}, nature \cite{claybornLivingRoomConservation2019}, road \cite{wangEducatingBicycleSafety2016}, sea \cite{l.calviVRGameTeach2017}, office \cite{steinfeldBeThereWhen2020}, truck \cite{korsBreathtakingJourneyDesign2016}, doctor's office \cite{deladismaMedicalStudentsRespond2007}, and home \cite{hamilton-giachritsisReducingRiskImproving2018}. A selection of different spatial contexts is presented in Figure~\ref{fig:spaces}.

\begin{figure}[b]
    \includegraphics[width=0.8\textwidth]{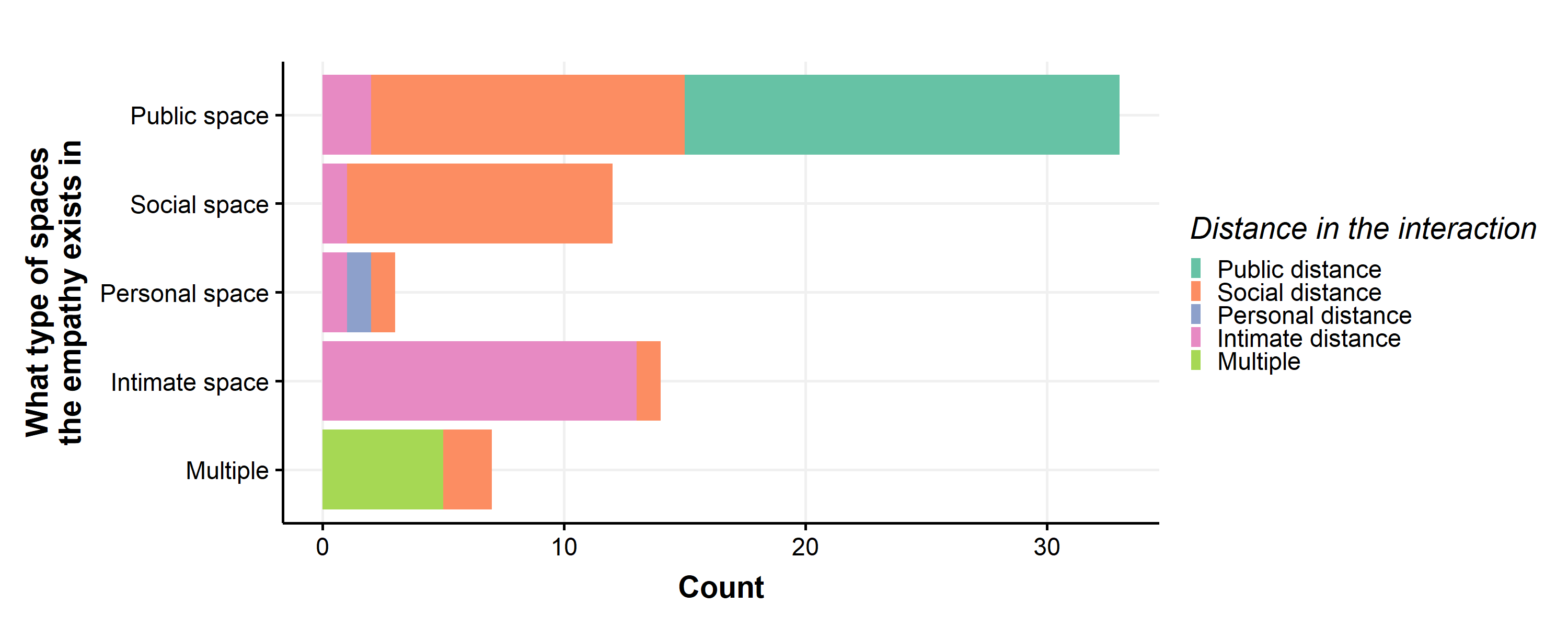}
    \centering
    \caption{Space of the empathic interaction compared to the distance.}
    \label{fig:space_vs_distance_interaction}
\end{figure}

Almost half of the papers (48\%) took place in a public space, as shown in Figure~\ref{fig:space_vs_distance_interaction}. Accordingly, most of the experiences in public spaces also used public distances. 20\% used intimate space, 17\% social space, 10\% multiple spaces, and 4\% personal space. 

\begin{figure}[b]
    \includegraphics[width=0.8\textwidth]{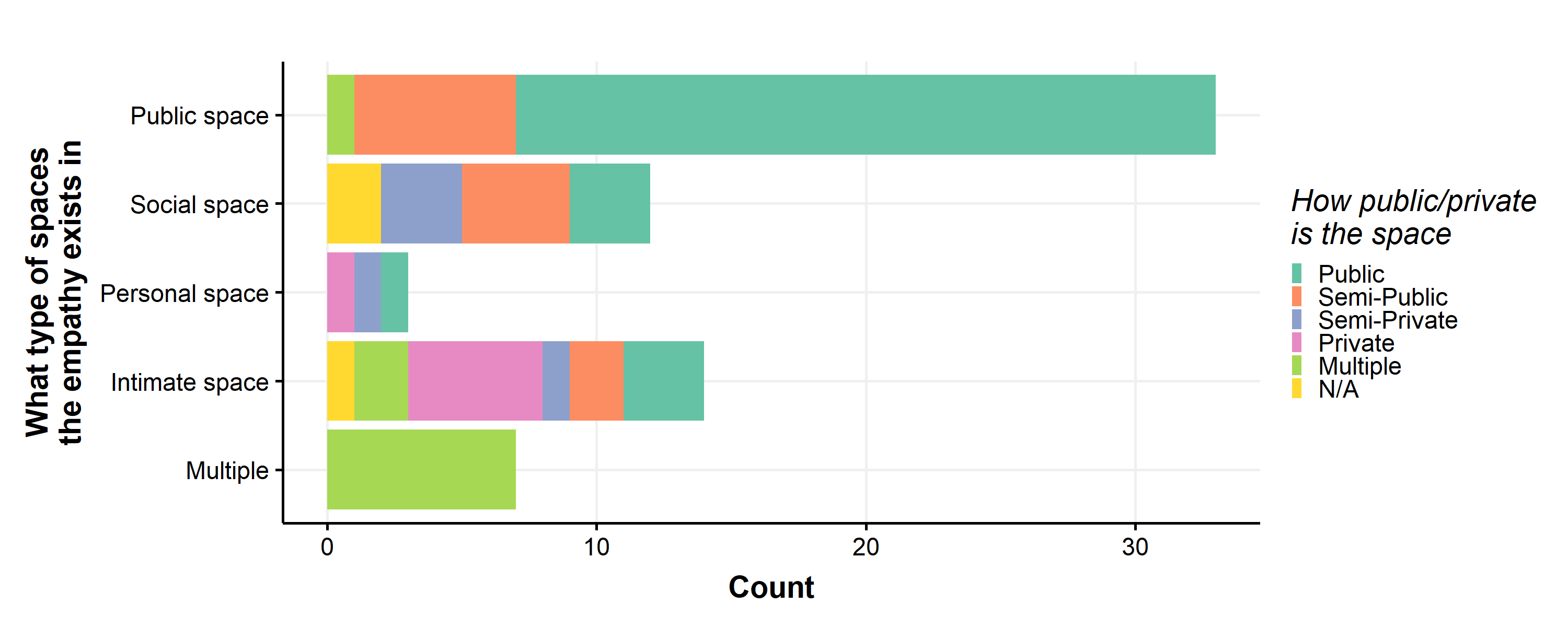}
    \centering
    \caption{What type of space is in question and how public or private are they.}
    \label{fig:type_of_space_vs_privacy}
\end{figure}

Nearly half of the papers (48\%) took place in a public context, compared to semi-public (17\%), multiple (14\%), private (9\%), and semi-private 5 (7\%) contexts, as shown in Figure~\ref{fig:type_of_space_vs_privacy}. In three papers, the privacy of the place could not be determined (N/A)~\cite{e.mavromihelakiCyberball3D3DSerious2014, ohVirtuallyOldEmbodied2016, guareseEmpathyPhantomLimb2020}. The private context of the home was used effectively in the study by \citet{hamilton-giachritsisReducingRiskImproving2018}. The study had sampled mothers for participants, who then embodied a young child and experienced aggressive behavior from their virtual mother, in the private context of the home.

We mark cases of direct sensory manipulation as intimate distance,  e.g., in the case of visual impairment \cite{p.r.jonesDegradedRealityUsing2018, g.m.rodriguezImmersiveApproachVisualizing2017, d.alexanderDesignVisualDeficit2020, werfelEmpathizingAudiovisualSense2016, s.misztalSimulatingIllnessExperiencing2020}, schizophrenia \cite{kalyanaramanVirtualDoppelgangerEffects2010}, or anxiety \cite{v.russellHAVEExperienceInvestigation2018}, due to the immediacy of experience. However, the decisive location of experiences, and by extension, distance from the person experiencing them, is difficult to assess. This aspect of experience does not translate well into the geometric understanding of space, and more vague descriptions can be more meaningful. In other words, proxemic and empathic distances are not easily understood together.


In most cases (61\%), there was a one-to-one relationship between the user and the \textit{Other}. It means the user was alone in the experiment and was focused on the experiences of a single entity. Then, 19 studies (28\%) used one-to-many relationships, 6 studies (9\%) used many-to-many relationships, and 2 studies (<2\%) used many-to-one relationships. To highlight one-to-many relationships, studies used a classroom bullying scenario through the eyes of the teacher or the person being bullied \cite{stavrouliaDesigningVirtualEnvironment2018}, experience of multiple occasions of sexual harassment \cite{venturaHowDoesIt2021}, and an educational approach to teach different religious practises \cite{johnsonUsingVirtualReality2018}. Many-to-many relationships focused on a more complex setup of users in an empathy situation. \citet{menzelEffectivenessPovertySimulation2014} studied poverty by placing the users in different family roles, each with their own  goals, to experience poverty and build empathy. In a meditative application, \citet{salminenBioadaptiveSocialVR2018} set up an experience where multiple users participated in synchronized breathing exercises to elicit empathy. In the Cuban Missile Crisis simulation by \citet{stoverSimulatingCubanMissile2007}, the students assumed roles of Cuban, Soviet, or USA officials during the Cold War, and developed empathy through understanding the complex relationships between the different countries, instead of a sole description of individuals, i.e., soldiers' faces with tension and fear. The two papers using many-to-one relationships employed multiple interaction possibilities to feel the autism of a single person \cite{haddodFeelAutismVR2019} and historical empathy where multiple students imagined life in a bomb shelter \cite{schaperDesignVirtualHeritage2017}.

\section{Discussion} \label{section:discussion}
As shown from the description of our included articles, there is clear variance in the use of different technologies, empathy approaches, and spatial contexts. Based on the findings in Section~\ref{section:synthesis}, we now tie these different threads together and raise an in-depth discussion on the different dimensions of designing empathy tools for XR.

\subsection{Spatio-Temporal Considerations for Fostering Empathy}
\paragraph{Understanding Spaces through Context}
Spatiality offers an analytical lens for examining empathy tools, as specific spatial contexts can be understood differently by different people. For instance, in the educational approach of presenting different religious sites and practices \cite{johnsonUsingVirtualReality2018}, the various meanings related to the temples, synagogues, etc., are not readily understandable by all people experiencing the environment. As such, the users of empathy tools are using their present understanding to make sense of the presented, possibly foreign, contexts.

To understand how these different modes of experiencing spaces work, we borrow the notion of ``\textit{context collapse}'' \cite{davis_context_2014}, which describes how all-encompassing social media makes it difficult to keep an individual's social circles separate. As people tend to adjust their behavior based on the social context, person may find it difficult to act meaningfully when these different social contexts are mixed. The context can blur either intentionally (``\textit{context collusion}''), or not (``\textit{context collision}''). \citet{davis_context_2014} recognize events such as weddings and funerals as examples of contextual collapses in the real world, where maintaining a singular role and cohesive behavior in a mixed group of friends, family, relatives, and acquaintances can be difficult. While the physical world is tied to the constraints of space and time, the virtual environments are more fluid by definition, enabling more contextual blur. We see use in approaching the continuum of virtual empathy (see Figure~\ref{fig:VC_continuum}) through contextual collapse, when the real and virtual contexts are intertwined. For instance, the spaces and norms that VR spaces represent in the real world can clash with the ``no rules'' context of VR \cite{nakamuraFeelingGoodFeeling2020}. The results of this context clash was seen in the VR metaverse application Horizon by Meta in 2022, where users faced sexual harassment due to a lack of personal boundary in the virtual space, leading to an implementation of an enforceable personal boundary to mitigate unwanted behavior\footnote{\url{https://gadgets360.com/social-networking/news/meta-horizon-worlds-vr-metaverse-personal-boundary-ring-feature-sexual-harassment-report-2750506}}. The lack of pre-existing context in virtual reality applications and the unlimited representations afforded by virtuality lead to a wide array of new contextual implementations, and contextual collapses. On the contrary, AR experiences are much more dictated by the spatial context of our physical surroundings. Considering the number of viral VR ``fail'' videos where immersed users forget their physical environment and hit themselves or others, it is interesting how similar examples of AR are hard to come by \cite{daoBadBreakdownsUseful2021}. As such, we posit the contextual collapse is greater with more immersive technologies, that is, VR. In order to truly experience empathy in virtual environments we suggest one needs to suspend their disbelief about the virtual experience and accept the new context it provides. 

In this sense, the users needs to give in to empathy in a virtual environment. While a person's characteristics affect the elicited empathy, the virtual environment and its context play a role. A believable XR experience builds a more understandable context and supports the user's willingness to empathize. We can also draw parallels between the privacy of space and context collision. In a private setting, regardless of the level of virtuality, the presence of strangers introduces contextual collision. On the flip-side, in public environments presence of other people can enable contextual collusion, leading to encounters missed in private settings. For instance, placing the user in a public environment, such as a busy metro, but focusing on private matters can bring a contextual collapse that seems unbelievable. Then again, unbelievability can act as a device through which foreign experiences are accepted. One study used the spatial context of a drug store, to highlight the disturbing personal symptoms of schizophrenia \cite{kalyanaramanVirtualDoppelgangerEffects2010}. The paper goes on to suggest that ``\textit{simulations without the context of \ldots supplemental resources may lead to outcomes that are not only inferior but are even counterproductive to learning goals}'', enforcing the need to use necessary contextual support in empathy tools.

\paragraph{Spatial Experience for Eliciting Empathy}
The discussion above highlights the contextual understanding of spaces for empathy, but we can approach spaces from another perspective as well. That is, how could a person's spatial experience elicit empathy? In other words, is the way someone experiences spaces something we can empathize about? What kinds of experiences could these be?

In this manner, we see two different sources for empathic spatial experiences. First, starting from the original concept of empathy, Einfühlung, we can see how art and architectural forms can be understood as a source of empathy \cite{galleseMotionEmotionEmpathy2007}. Recent advances in neuroscience have shown links between the visual perception of art and the mirror neuron system \cite{schottPicturesPainTheir2015}. However, this approach requires a distinction between empathy and affect. An impressive architecture can elicit strong feelings, but strictly speaking, it is not empathy however valuable as an aesthetic experience it might be. Regardless, we see value in the virtual environments, where the structural forms of the environments are not bound to the rules of the physical world, which enables experimentation with new kinds of forms and empathic experiences.

The second way the spatial experience can be emphatized is through someone else's experience of space. As empathy tools use experiences to elicit empathy, we see spatial experiences as another avenue for empathizing. These spatial experiences can include place-dependent experiences, such as safety, accessibility, or aesthetic appreciation. As our understanding and experience of spaces is multilayered and complex and tied to personal history and sensemaking, the richness of experience could serve well to elicit empathy. However, a concentrated study is needed to clarify how we understand and experience these shared meanings of space. Similarly, we could benefit from further explorations on how a given empathy tool works in different spaces. For example, how could a person's experience of social anxiety differ in public and private spaces? Or how the use of multiple POVs could show nuanced aspects of the spatial experience? Furthermore, real-world spaces are living and the atmosphere can change in an instant, for example in an accident, act of terrorism, or fire. Introducing the layer of empathy and virtuality raises the question of how they can address and convey changes in the atmosphere of the place? Moreover, in an AR system, the user is collocated in the space with others. When the empathy tool elicits the experience of others successfully, how does it change the atmosphere? Through these explorations, we see architecture and HCI form a more nuanced understanding of spaces.


\paragraph{Frictional Spaces}
The XR spaces aimed at eliciting empathy are, in essence, environments for learning. What is unique in these spaces is that they provide connections between humans, and other humans, animals, nature, or objects.

As the goal of the empathy tool designs is often not the artifact in itself, but rather, the experience of empathy, its methods resemble alternative design approaches. The concept of ``frictional design'' \cite{pierceTensionProgressionGrasping2021} describes how alternative designs have goals that differ from traditional, progressional, concepts of designs. Following this notion, ``\textit{frictional spaces}'' can be understood as spaces with an alternative purpose. Dwelling, described as the act of making one's home \cite{heideggerBuildingDwellingThinking1971}, works towards building immunity from the outside's unpredictabilities. As such, dwelling is orthogonal to frictional spaces. In contrast, the frictional spaces are not meant as environments for continuous dwelling, but rather, they are experienced similarly as one experiences art.

Following this line of thinking, could one consider the real and day-to-day spaces we share with others as frictional spaces? We do not always prefer coexisting within a space, but it can lead to chance encounters, through contextual collusion. Using the examples of a bus stop and a spatial art installation we highlight the range of frictional spaces from banal to extraordinary. As such, our everyday spaces can indeed be considered frictional spaces. What is emblematic of these frictional environments is that inhabiting them is not necessarily comfortable, or even the end goal. Transitional and liminal spaces also share these qualities but are more defined by their use of moving through or waiting in, for instance.

The virtual spaces offer a simulation of these frictional spaces. Notably, virtuality also leads to a lack of real-world consequences. This might explain the challenges in using virtual environments to elicit real-world empathy. \citet{nakamuraFeelingGoodFeeling2020} argued all spaces in VR are public as there is no privacy to maintain. The non-playable actors in the virtual environment are not in control of the public and private spaces of the environment unless specifically designed. In contrast, AR technologies are grounded in the real spaces and their concept of privacy. Similarly, XR (AR and VR) companies also need to be considerate for protecting users' privacy. For instance, the sensor-heavy XR technologies can gather data from a wide range of human actions and their surroundings, e.g., their behavior, biometrics, speech profile, and other people you share a space with\footnote{\url{https://www.eff.org/deeplinks/2020/08/if-privacy-dies-vr-it-dies-real-life}}.




\paragraph{Empathizing the Past, Present, and Future}
With XR technologies, empathy tools are not limited to contemporary experiences. If there is a perspective known to a sufficient degree, it can be attempted to empathize. Now, the design of these empathic experiences relies on assuming some parts of the experiences of the \textit{Other}, like all empathy simulations, so the development process should be carefully valued between accuracy and the effect. For instance, take the world of literary fiction, where a skillful writer can convey a host of feelings to the reader, to elicit empathy towards the characters. As an example, a short story ``\textit{Land Deal}'' by  \citeauthor{murnaneCollectedShortFiction2018} examines the relationship between Australian aboriginals and colonizers through an exchange of goods for land. The tensioned dynamic between these two parties resolves in the aboriginals' suspension of reality, and in the belief that they are living a dream. The story is a display of writing to highlight the experiences of the \textit{Other}.

Following a narrative approach, the studies on historical empathy found in our survey serve as a glimpse to the past \cite{sweeneyUsingAugmentedReality2018a, stoverSimulatingCubanMissile2007}. Using historical knowledge, we can present viewpoints that would be otherwise inaccessible to us. These are most often based on some understanding of the past, e.g. through written records, historical knowledge, and more. However, we see that historical fiction could be a new avenue to explore, as it was not found in the empathy tools in the present survey. In empathizing the past, we can enrich history and bring more nuanced perspectives into our present day.

The benefit of empathizing the present allows the users to relate their experiences more easily, as there is most often more contextual common ground. As empathy tools often aim to foster a behavior change, it could be argued that empathizing the present can be more valuable for the current times. For instance, empathizing hardships of wartime, when living in peace, might not be a directly applicable experience. 

Empathy can also be used to target future speculations, although such aspect was not found in the articles considered in our survey. However, \citet{thibaultTransurbanismSmartCities2020} used pastiche scenarios to highlight different aspects of the future of smart cities. These scenarios can elicit empathy and provide viewpoints that are not limited to experiences that have happened. The literary genre of science fiction is often attributed to future speculation and can provide these views of the future that elicit empathy. These alternative futures can shape our present understanding, and widen our possibilities for the current times. As such, the future prospects offer a non-linear mode of thinking and exploring our present world.

We can also focus on how long we experience the empathic stimulus. Anecdotally, nearly all empathy tools are relatively short experiences (under 30 minutes), but research on longer immersion (e.g. in metaverse) is close to none. However this can be partially explained as VR technologies can be prone to cause sickness, their use is often limited to shorter durations, especially for the uninitiated. In contrast, AR environments can be a more ecological approach for eliciting empathy in longer durations due to a smaller amount of sensory manipulation and therefore less sickness. Contrasting these XR technologies with literature, we see that most literature is experienced over a longer period of time, which possibly serves to build a stronger feeling of empathy. However, research on the long-term exposure to the empathy stimulus is limited, and further research would help to elucidate how the duration of the experience affects empathy. 


\subsection{Facilitating Empathy in XR and Metaverse}
To elicit empathy in a person, the choice and design of the suitable stimulus play a large role. The shape of the empathy tool can be dictated by the desired experience, available technologies, and resources, or the context of the empathy situation. Accordingly, these aspects can act as the driving forces behind developing the empathic stimuli and facilitating XR empathic experiences.

\paragraph{Empathy Technologies in the Virtual-Physical Spectrum}
In the studies that compared the empathic effect across different technologies (e.g., \cite{bujicEmpathyMachineHow2020a, calvertImpactImmersingUniversity2020, steinfeldBeThereWhen2020}, the more immersive approaches often elicited more empathy, as opposed to low immersion conditions such as reading or viewing the experience on a desktop monitor. Similarly, another study expressed this immersion through a model where immersion and related concepts (e.g., presence, flow) are linked with more empathy \cite{shinEmpathyEmbodiedExperience2018}. While evaluating several technologies is a sensible choice in terms of a study design, it raises the question, to what extent are these experiences are comparable? For example, comparing a reading task to a virtual HMD experience in terms of immersion (e.g., \cite{herreraBuildingLongtermEmpathy2018}), naturally the modern understanding of virtual reality immersion is greater in the latter. Understanding the nuanced way these different technologies operate supports designing the empathy stimulus more meaningfully. Furthermore, the same holds true when using different yet complex technologies. In augmenting the physical world with digital content, AR experiences can strengthen empathy in environments familiar to the person through additional information, whereas VR experiences support experiencing empathy in contexts which would be otherwise not "found" in person's own life. It is worthwhile to mention that, albeit this paper has traced back to earlier works in 2006, we rarely find CAVE systems for eliciting empathy, which only appeared in 2013, 2019, and 2021. The CAVE systems refer to some room-size immersive environments, which are primarily employed in industrial applications, such as staff training and consumer product experiences~\cite{Muhanna2015VirtualRA}.

The 360 VR videos proved to be an ecological choice for many (12\%) study designs (\cite{v.camilleriWalkingSmallShoes2017, raduUsing360VideoVirtual2021,tanEmbodyingSenescencePerforming2019,munteanCommunicatingSustainableConsumption2020, venturaHowDoesIt2021, nelsonVirtualRealityTool2020, johnsonUsingVirtualReality2018, griffinExploringVirtualReality}). We were surprised by the relative prevalence of VR 360 videos, which in the hindsight are relatively simple to use. The rise of VR as a medium can be explained through the increased adoption of consumer-grade VR headsets, reaching a critical mass~\cite{Kelly2021WhoAV}.  Also, the existence of 360 video platforms, such as YouTube, helps facilitate highly diverse empathy experiences. Even more, the use of playlists allows for a curated selection of empathic material, such as in the case of religious education \cite{johnsonUsingVirtualReality2018}. If one considers the social benefit of using an empathy tool, an easy-to-use platform with curated experiences can support eliciting empathy on a wider scale.   From our observation of various empathic scenes in VR and 360 videos, virtual environments have the advantageous properties of malleability and immersiveness that facilitate various imaginary scenes. It is important to note that certain scenes, e.g., child abuse \cite{hamilton-giachritsisReducingRiskImproving2018} or drug usage \cite{christofiVirtualRealitySimulation2020}, cannot be replicated in real-life scenarios, due to the high costs of human actors and ethical issues (e.g., mental and physical damages caused by killing and hurting) of replicating such scenes. Thus, virtual environments play an indispensable role in eliciting empathy through putting users in scenes with strong sense of presence~\cite{Grassini2020TheUO}. Nevertheless, prior works reveal that the skills or knowledge acquired in virtual environments can be disconnected from the scenarios in physical world. Therefore, augmented reality can play an auxiliary role in visual cues to remind the acquired skills and knowledge during applicable contexts~\cite{Cooper2021TransferOT}. Similarly, AR can potentially complement VR to extend the entire cycle of empathic journey -- VR aims to elicit users' empathy, while AR reminds users of the experienced empathic scenes in matched contexts, achieving transfers of ``\textit{empathic} learning''. Even though we only examine limited works leveraging AR, understanding how social, location-based games such as Pokemon Go \cite{allowayGottaCatchEm2021} can elicit empathy or historical empathy for enhanced learning \cite{efstathiouInquirybasedAugmentedReality2018}, AR can bring in-context experience and arouse empathy in physical spaces. 


\paragraph{Interaction Modalities}
Virtual entities can merge with our physical living space and provide various functionalities and utilities~\cite{Lee2022TowardsAR}. Visual overlays in augmented reality can serve as an effective visualization tool to inform users regarding e.g., privacy threats, along with delivering empathic content to users. Thus, virtual spaces could serve as a more explicit and swift communication than traditional 2D UIs~\cite{para-icmi}. 

On the other hand, researchers in the broadest domain of virtual environments (i.e., not limited to empathy-related studies) attempted to explore multiple interaction modalities to enrich the user experience, e.g., gestural interaction plus haptic feedback~\cite{hexa-configuration}. In such empathic immersive environments, the users had used one or more interaction modalities, including tangible devices like keyboards and joysticks, as well as embodied interaction like gestures and speech (Figure~\ref{fig:used_technology_by_interaction_type_by_locomotion}). As stated in prior work on city-wide user interaction with immersive environments~\cite{Lee2022TowardsAR}, user bandwidths, i.e., the user ability to convert their intents into actions in virtual environments, drops significantly in virtual environments.

When both the interaction modalities and the levels of empathic virtual spaces are considered together, two questions remain unexplored -- 
(1) What is the threshold of user bandwidth that guarantees the experiences of eliciting empathy ;
(2) Do richer interaction modalities also elicit higher levels of empathy?


\paragraph{User Agency in Experiencing Empathy Tools}
Our collected articles reveal that empathy can be taught (i.e., aroused) in virtual environments. For instance, examples include the training of staff in medical services and education institutes \cite{kleinsmithUnderstandingEmpathyTraining2015, tettegahCanVirtualReality2006}. Virtual environments allow the staff to be situated in the role of victims, e.g., physical or verbal violence. Through such perspective-taking approaches, the staff had gained awareness of violent behaviors and understood the perspectives, and perhaps experiences, of the persons being abused. 

Noticeably, the content designers of virtual environments bring a level of \textit{agency} to their audiences. In virtual environments, \textit{agency} refers to the extent to which users can interact and modify objects in the scenes \cite{jicolEffectsEmotionAgency2021}. Users with virtual environments of high agency levels can employ a more active approach when interacting with the empathic content to experience empathy. Therefore, eliciting empathy is subject to agency, creating either \textit{curated} virtual experiences \cite{munteanCommunicatingSustainableConsumption2020} with low agency or \textit{free-to-wander} experiences \cite{helgertStopCatcallingVirtual2021} with high agency. 

Apart from the agency that describes the flexibility and richness of virtuality, we consider that locomotion (Figure~\ref{fig:used_technology_by_interaction_type_by_locomotion} is a critical user-centric factor that empowers agency. Locomotion, through various interaction types, enables greater agency for the user in the virtual environment. Users leverage interaction techniques (gestures and keyboards) to explore (e.g., walk and search) the environments and the intended experiences regarding empathy. 

For instance, the underwater diving simulation \cite{l.calviVRGameTeach2017} allows the users to freely explore a seascape in VR. The user can control a virtual Diver Propulsion Vehicle (DPV), to locomote underwater. The game focuses on exploring the diverse seabed scenes to elicit empathy towards underwater sustainability. As a side-effect, the authors note the safe VR experience can be used to surpass users' phobias and physical disabilities. 

As the collected articles employ highly diverse scenarios, but inconsistent evaluation methods, we cannot state to a definitive conclusion on how agency and users' active participation correlates to the effectiveness of eliciting empathy. Nevertheless, it is important to note that the learning process of empathy can be facilitated by a process of active participation (see e.g., \cite{stoverSimulatingCubanMissile2007}), supported by a high agency level. In contrast, a low agency level only allows users to receive information in less interactive yet unidirectional flows. Therefore, the relationship between the agency and the easiness of experiencing empathy requires further investigation. A very limited number of studies reflect the accuracy of user experience in such virtual empathy situations, and we have very limited knowledge of how faithful to the original experience is the simulated experience. In addition, as the existing works have no comparable virtual scenes and different empathy goals, the connection between designing fictional events and the process of empathizing with users' personal experiences poses significant research opportunities.



\paragraph{Embodiment and Empathy}
The included papers represent embodiment beyond humans and include animals, nature, and more. As such, they provide a perspective into non-human entities as well. As no paper in our survey used multiple types of embodiments, we highlight this as a future research potential for studying empathy. To this end, we present two relevant explorations. First, \citet{krekhovHumanAnimalsEscape2019} studied how non-human representations can be used in VR. Their experiment with three animals (rhino, scorpion, and bird) implies an illusion of virtual body ownership (IVBO). Through the experiments with different animals, they recognize more work is needed in both hardware and software to produce believable and enjoyable VR experiences with animals. Second, the experimental video game ``\textit{Everything}''\footnote{\url{https://www.davidoreilly.com/everything}} by artist David O'Reilly empowers the player to embody numerous amounts of non-human entities. The player can embody and control things as small as particles, systems as large as galaxies, and everything in between. In this process, the game focuses on the interconnected nature of the universe and builds perspective-taking on a massive scale. The game serves as a useful frame of thinking in multiple scales and contexts for empathy. 

To this end, XR serves as a fruitful enabler of experiences with varying embodiments. Through VR, we can craft new experiences, which are less related to the spatial context of the physical world, and are more focused on the unique possibilities of world-making and experimentation with scale. As shown in MR studies in our included articles, the additional sensor and interaction technologies allow a more immersive and embodied empathy \cite{leeWhoWantsBe2019}. As the AR technologies operate in the spatial context of the real world, its inclusion of non-human empathy can help to ground us in our environments. The AR enjoys a strong potential to present augmented information in its context. In this sense, while in our view AR is often related to human embodiment, it can show the world through adjusting the perception of real-life (e.g., see \cite{kellyOurDogProbably2020}).



\subsection{Empathy at Large: Limitations, Criticism, and Ethics}
Finally, we can consider empathy beyond the scope of a single study. We understand and use empathy through different means and with different goals in mind. Who designs these empathic experiences matters and what values are represented therein. Through careful consideration, we evaluate empathy and its virtual possibilities as a fair mechanism for self-reflection, for the society at large. 

Developing empathy tools raises the question: who designs them and with whose experiences? As empathy experiences are most often concerned with personal viewpoints, using someone's experiences for empathic stimulus is challenging. The level of fiction in the  representation of the emphatized experience affects how close to the ``source material'' one is. In some cases, representing someone's experiences can be understood as a process of sustaining stereotypes \cite{nakamuraFeelingGoodFeeling2020}. For instance, while racist experiences can be used to elicit empathy towards their victims, they also reproduce those events. As such, being considerate and respectful of someone's experiences is essential, and a critical approach is needed to produce empathy that serves the common good. While empathy is most often seen as a positive aspect of human social behavior, it is the methods with which we produce and reproduce empathy, that can also harm. Another concerning aspect we found in our surveys as well as the varying respect of privacy. Filming vulnerable populations such as children, underrepresented, low-income, homeless, etc., has an inherent risk of exploitation. Respecting others' experiences also requires respecting their privacy and choice not to be used as means for empathy. To engage with the people of these experiences, mutually informed participatory design methods could help produce more sustainable empathy tools.


We can consider empathy research taking place in the public and private sectors. Naturally, academic research in the public sector aims to understand how empathy works with regard to technology and digital media. The works covered in the present survey are most often examples of public research. Moving towards the private end of the spectrum, there are various entities interested in empathic experiences. For instance, if we consider empathy tools in the wide sense of digital media, we can see how the video game industry is concerned with eliciting empathy. Character and narrative-driven gamified experiences provide various human perspectives with different levels of realism. The many visions of metaverse align with these character-driven experiences, with the distinction that there is no scripted narrative structure. The incentives of the company, institute, or entity are relevant as they shape the goals for the empathic experience.

Looking from the perspective of the users reveals another point of discussion: for what purposes do people use the empathy tools? In the realm of entertainment media, we can see how empathic games can provide engaging entertainment. Then again, serious games have goals less related to entertainment \cite{deterdingGameDesignElements}, and more related to education and perspective inducement. As only a few empathy tools presented here are purely entertaining by nature, we see designing these approaches to have more similarities with serious games. The goals of serious games are in a way directed outward, towards other people. As such, we can see empathy through the lens of doing social good, through nuanced behavioral change. In contrast to the financial care towards others in social good like charity, behavioral adjustment with empathy tools could serve as another form of tool-assisted altruism. To see empathy as the social glue also requires us to recognize moral aspects of understanding another, and ultimately, how should person's character be developed. Granted, this line of thinking can lead to wicked problems about the way person's behavior should be changed, and who has a say in the matter.


\begin{figure}
    \centering
    \includegraphics[width=0.35\textwidth]{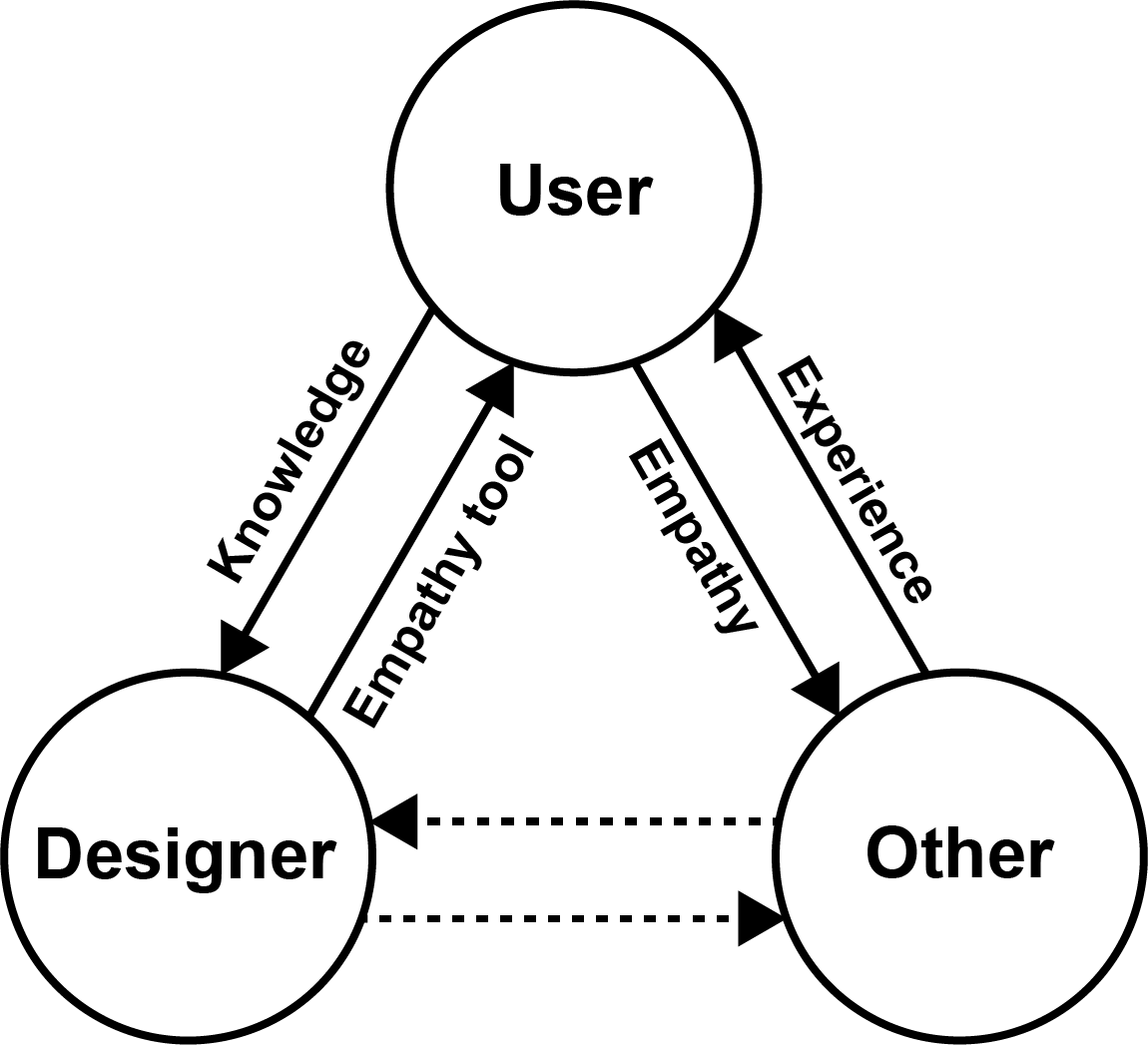}
    \caption{Stakeholders and the resulting outcomes in the empathy tool design process. The dashed line represents a relationship that is not readily understandable but is nonetheless important for sustainable empathy.}
    \label{fig:empathy_relationship_chart}
\end{figure}

It is also relevant to consider what relationships are fostered, and what understanding is transmitted when designing and using empathy tools. The designer, the user of the empathy tool, and the person whose experiences are used for empathy are in an interesting dynamic of meaning-making. We can represent these relationships diagrammatically through Figure~\ref{fig:empathy_relationship_chart}. Notably, the relationship between the empathy tool designer and the \textit{Other} are not easily explainable. To what extent does the designer participate with the \textit{Other} in designing the experience? Conversely, how does the \textit{Other} benefit from this relationship? Having these multiple lenses means understanding empathy purely from a single viewpoint in a robust fashion can be difficult. For instance, the designer expects and hopes to elicit empathy, which can inadvertently affect the user's experiences, which can be challenging in the academic context. Further, the person whose experiences are used might be compensated for their input but is never actually empathized in the real life. Similarly, the user can experience great empathy, while they might never encounter a similar situation in the future. This echoes \citeauthor{batesonSmallArcsLarger2016}'s statement how in the academic realm we often remove participants from their context but rarely place them back in \cite{batesonSmallArcsLarger2016}. To this end, we need to take care in how we respect the various stakeholders and the contexts they live in.

Often empathy tools are one-off interventions. Accordingly, like with other one-off interventions, there is a lack of engagement with cultural systems in a deep enough way. To produce a lasting impact, there needs to be a culture around eliciting and fostering empathy. It is this systemic way of approaching empathy that leads us to consider, how we can tie the experience of empathy into the fabric of everyday life? Introducing new technologies can have surprising effects, and there is no deterministic foresight to know what approaches work. The way we see it now, the empathy interventions are often quite separate and limited to individual instances, however successful they are in the scope of single research. These research efforts are by no means undermined, as their advances produce understanding towards a more empathic future. However, longitudinal and persistent empathy applications would clarify the role of digital media in eliciting empathy for real-world use. This should be done with care towards everyday life, to produce meaningful and adoptable results. It is this balance of tending and working with the existing systems of human life and culture while developing new technologies that serve the global and local good. Underlining empathy through the means of technology serves to connect us and mend what needs mending. As expressed by computer scientist Sep Kamvar, ``\textit{it is important \ldots to tell the stories that, when they become realities, will help to heal society}'' \footnote{\url{https://farmerandfarmer.org/succession/syntax.html}}. 



\section{Research Roadmap: Three Opportunities for Empathy Across Physical and Virtual Spaces} \label{section:research_roadmap}

After analyzing and discussing the existing literature on empathy and virtual environments in Sections 3 and 4, we propose a research roadmap and the corresponding open questions to support empathy research in XR.

\subsection{Designing and Researching Empathy with XR}
The prevalent trends in XR media indicate potential for fostering empathy. On top of the technologies supporting virtual environments, we need to thoroughly examine the scope and effects of XR-driven empathy. 
A key challenge is to provide an experience that successfully elicits empathy and, if possible, inspires taking action. Shifting from recognizing the experiences of the \textit{Other}, towards more action-driven outcomes serves to extend benefits beyond the simulation in the virtual environment. The articles in the present survey did not find meaningful differences in long-term empathy, and to this end, more work is needed to ascertain what factors could support empathy over a longer time period. Moreover, the  methodologies deployed in the studies are not useful in comparing the effectiveness between different approaches. Further research into the various stakeholders in empathy tool design could serve to promote more sustainable approaches for eliciting empathy. To this end, constructing a framework for developing and measuring empathy tool experiences with XR technologies would support proper practices, and help create new ones. Accordingly, the designers of empathy experiences have a great responsibility. For instance, a collaboration between social scientists, psychologists, anthropologists, architects, and designers is needed to develop a meaningful, responsible, and sustainable empathy.

\paragraph{Open Question 1: What methods support long-term empathy?}
More knowledge is needed on producing a lasting impact with empathy. As our findings show (see Figure~\ref{fig:empathy_measurement_by_longer_term_measurement}), the long-term impact of empathy intervention was measured only rarely, and usage of the empathy tools was often limited to one-off instances. To support building long-term empathy, we need a more systematic approach to facilitating empathy, and the empathic reality approach described below is a step in that direction. How the empathy experiments can work over time is ripe for more longitudinal work.

\paragraph{Open Question 2: How to foster non-human empathy?}
Empathy is mostly focused on the experiences of fellow humans (see Figure~\ref{fig:embodying_by_target}). With the rising focus on Animal-Computer Interaction (ACI) in computing (see, e.g., \cite{manciniAnimalcomputerInteractionManifesto2011}) and more-than-human approaches \cite{fieuwMorethanHumanApproachSmart2022}, we recommend a wider understanding of empathy for a more sustainable future. The direction in which we develop technology, and by extension our environments, should work with the multi-perspective lens of flora, fauna, and humans. In essence this approach encourages diversity through the cultivation of multiple perspectives, and helps to promote actions towards environmental challenges such as the climate change and sustainable behaviors \cite{brownEmpathyPlaceIdentity2019}. As shown by the limited count of non-human empathy, we promote the challenge of facilitating empathy beyond our species. While the reciprocal nature of empathy is difficult to ascertain, we suggest the empathic position towards flora and fauna builds awareness of co-existence.

\subsection{Re-Imagining Experienced Spaces: Temporal and Spatial Empathy} 
Spatiality is a beneficial lens for analyzing empathy. However, current research does not consider how the space affects empathy, and conversely, what possibilities empathy has for spatiality. To this end, we recognize two research opportunities. 

First, more empirical knowledge is needed on how spaces modulate empathy. As XR technologies have a wide range of input modalities and immersive capabilities, generalizing spatial experience is difficult. However, we suggest that proposing technology-specific guidelines for developing with spatial experience in mind would lead to actionable results. Further, with empathic content that is available and accessible in the spectrum of virtuality, we can open up new application areas for research in terms of spaces. For instance, how could these traditionally one-off interventions be installed in a physical space in the city? Alternatively, what kind of empathy could be fostered in the privacy of one's home? As such, the space and its context play into building a more nuanced understanding of empathy. As empathy tools are focused on immersing the user, we could see them serve a similar purpose as movie theaters or libraries. Further, the historical or futuristic forms of empathy are applicable in the educational contexts of museums and schools. New applications for empathy are in line with trends such as the empathic city \cite{biloriaSmartEmpathicCities2021} that have future-oriented strategies that move from the technocratic focus of smart cities into a more human centric focus. To this end, contextual collapse allows us to recognize how the contexts of physical and virtual spaces are brought together. This leads to a great challenge in the research of empathy tools. How are these different contexts interpreted, and re-interpreted? The designer of the empathy tool uses the experience of the \textit{Other}, and reframes it in a technological tool, using the designer's understanding of the context. Then, the user of the tool experiences using their contextual frame, and hopefully experiences empathy in this process. 




Second, we see potential in examining how spatial experience can be used to elicit empathy. A person's experience of space with regards to aspects like safety, accessibility, aesthetic experience, personal meanings, and social interactions can be fruitful triggers for empathy. This \textit{spatial empathy} serves to form empathic relationships through the experience of space. As spatial experience is related to an individual's sensemaking, there is great potential for communicating and empathizing these personal experiences of space. We see a few relevant focus for research into spatial empathy. Clarifying the relevant aspects of the spatial experience for eliciting spatial empathy highlights the potential for perspective-taking. Also, researching methods for spatial empathy would build understanding towards more effective empathy studies. For example, \citet{heissEmpathySpace2006} developed a prototype named ``\textit{Empathy Vest}'', which is an interactive vest that connects the user to the remote environment of someone else, through multiple sensors. We see this somatic approach of empathy as fruitful for eliciting spatial empathy, due to the holistic way we experience spaces, alongside cognitive and affective forms of empathy. Further, like the included articles in our survey, spatial empathy is not limited to contemporary experiences. Memory and data help to empathize the past, and speculations can focus on the future. Besides the linear understanding of time, we can follow \citeauthor{lefebvreRhythmanalysisSpaceTime2004}'s method, by approaching space through its temporal cycles and rhythms \cite{lefebvreRhythmanalysisSpaceTime2004}. What hourly, daily, monthly, seasonal, or yearly experiences of space are there? What rhythms are not linearly understandable or measurable? Most importantly, how can these nuanced aspects of time help to evoke spatial empathy? Spatiality and temporal experiences provide challenges as well as the potential for eliciting empathy.

\paragraph{Open Question 3: How can spatiality be better understood in XR?}
The present understanding of spatial experience in everyday physical environments is not directly applicable to the wide range of extended reality technologies. The current research in VR immersion etc. is partially related to spatial experience, but more focused on the believability of the virtual experience, and what design aspects are most profitable. As such, a successful technical implementation leads to a more immersive experience, enabling further spatiality. Different virtual applications are still quite implementation-specific. Further, as different aspects of spatial experience, and especially perception, have been studied in virtual reality (e.g., scale and gravity \cite{pouke_plausibility_2021} and distance \cite{maglioSpatialOrientationShrinks2014}) more information is needed on how this knowledge is applicable in real-world environments. In the range of XR, the level of virtuality challenges developing an all-encompassing theory of spatiality. To this end, we posit that exploration into different spatial experiences in XR is conducive to building more impactful and believable empathy tools.

\paragraph{Open Question 4. What tools can support interaction designers prototyping temporal and spatial empathy?}

Both temporal and spatial elements can significantly impact the XR experiences and the delivery of empathic spaces. Nevertheless, the required efforts for blending our spaces are high, primarily due to the readiness of XR prototyping tools. In brief, the XR user experience is not yet well studied by interaction designers, while the development time and the lead time of collecting feedback are long. As reflected in an interview of 26 AR/VR designers and developers~\cite{CollaborativeXRSurvey}, the current processes of XR developments cannot effectively bring synergy between the technical AR/VR developers and content creators, while presenting limited communication channels to collect feedback from end-users. Researchers have proposed various prototyping tools to address the issue, such as  \textit{SpatialProto}~\cite{SpatialProto}, but none of the existing works, to the best of our knowledge, serves  prototyping XR in empathic contexts specifically. Remarkably, the knowledge of collaborative interactions (many-to-many) and user agency also requires further research (Sections~\ref{section:synthesis} and~\ref{section:discussion}), and we call for diversified tools to promote the understanding of users' empathy in XR, especially when designing interaction for empathic reality requires doubled efforts compared to usual XR applications, for the investigation of both users and the \textit{Other}. 


\subsection{From Virtual Environments to the Metaverse}
From the exhaustive list of studies considered in this paper (Appendix~\ref{appendix:included_publications}), we see significant research efforts on pursuing empathy development in people through XR technologies. The collected works demonstrate empathic content going beyond personal, temporal, cultural, and geographical limitations. Empathic content can be presented through virtual environments that are characterized by plasticity, expansibility, and scalability. Virtual environments enable the users to understand the perspectives of others by stepping into one's shoes. As such, it results in enhanced empathy and a better understanding of others' sufferings. Users with virtual headsets can gain a sense of presence and realism~\cite{Grassini2020TheUO}, in which a first-person view can bring the users an illusion of experiencing someone else's suffering. Still, such empathy can be limited to solely immersive worlds with VR headsets. We acknowledge that the rising availability of consumer-grade mobile VR headsets also gives a strong incentive to attract research attention on VR-driven empathy. Meanwhile, limited options and high prices of AR headsets have led to limited interests in facilitating empathy with virtual-physical blended environments (AR). 

It is also worth mentioning that the current interest in the metaverse will accommodate multitudinous new content and numberless content consumers. As stated in~\cite{Lee2021AllON}, the metaverse situates at the stage between \textit{Digital Twins} and \textit{Digital Natives}. The ultimate phase of the \textit{co-existence of physical objects and virtual entities} projects that the metaverse eventually will become ubiquitous in our daily life through a highly blended virtual-physical reality. Though the current metaverse, defined by mostly technology giants (e.g., \textit{Meta} and \textit{Microsoft}), refers to sole virtual environments, we foresee that the metaverse will result in inclusive environments and simultaneously provide building blocks for building the metaverse towards empathic dimensions. With very limited knowledge of AR-driven empathy, the metaverse will hardly reach the ultimate phase. 
To this end, it is necessary to link the sole virtual environments with virtual-physical (blended) environments~\cite{m2a-lam, a2w-james}. As a prior study revealed that the combination of AR and VR can further engage the users~\cite{Cooper2021TransferOT}, more investigation should be done to achieve the potential of the empathic metaverse.

As the majority of studies only reveal a snapshot of enhanced empathy after the participants had experienced the empathic virtual environments, we cannot give a definitive conclusion on what will be the exact role(s) and limitations of such empathic virtual environments in the metaverse. However, our observations among the collected articles give obvious cues that the virtual environments have strong efficacy in cultivating care towards others, e.g., the environment \cite{nelsonVirtualRealityTool2020}. Albeit we do not have a piece of solid evidence showing that virtual environments give long-term empathy to the users, immersive technologies could support the very first step of building up a harmonic and mutual-understanding society, through inclusive- or empathy education. When metaverse technologies can support such scenarios blended into our physical world, our immersive urban environments should own a capacity of reflecting multiple, and perhaps conflicting, interests of various stakeholders, as well as their difficulties in a shared and open community~\cite{Lee2022TowardsAR}.

\paragraph{Open Question 5. What possibilities exist for planet-scale virtual-physical blended environments for empathic reality?}

The empathic experiences in immersive virtual environments described in this paper have been constrained by the state of immersive technology. As such, these experiences are confined to small-scale virtual and physical spaces, with a single user at the core of the experience. Besides, many of these experiences focus on fully virtual worlds, with little physicality. Recent advances in extended reality technology are enabling many larger-scale experiences, whether in terms of physical space, digital space, or several concurrent users. Planet-scale virtual and mixed environments are under concrete considerations through proposals such as the metaverse. Such a shift in technology is likely to affect the design of empathic experiences. Enabling planet-scale (digital or mixed physical-digital) experiences shared between a multitude of users enables greater flexibility in the environments and actors, multiplying the possibilities for empathy development. 

\paragraph{Open Question 6. What are the vital roles of autonomous agents when `empathic' reality goes wrong?}
Besides extended reality, advances in artificial intelligence enable the creation of digital agents that can dynamically adapt to the user's behavior, to guide the experience and ensure accomplishing the primary goals. We foresee the future of empathic 
experiences to be more natural and open-ended, relying on the interaction between multiple simultaneous users and digital agents. However, open-ended experiences raise the question of accomplishing the primary empathy goals. An empathic experience where most actors are players may rapidly turn into a game where users collaborate towards turning an unpleasant experience into entertainment. For instance, a simulation of war conditions may turn into a collaborative survival experience which would negate certain of the intended aspects such as distress or loneliness. At the other end of the spectrum, it is necessary to address phenomenons such as cyber-bullying, which may further distress users placed into uncomfortable experiences. A delicate balance will have to be achieved, where the simulation adapts to users' behaviors to avoid unforeseen actions, and perhaps autonomous agents characterized by stewardship and trustworthiness can audit varying empathic experiences in many virtual worlds in the metaverse era.

\section{Conclusion} \label{section:conclusion}
This article surveys the area of empathy-driven immersive extended reality (XR) technologies, analyzing different ways by which virtual environments can simulate and trigger empathy. We especially focus on the spatiality and the contexts that these empathy tools take place in. Consequently, we bring an understanding of how users can enhance their awareness and empathy towards others' experiences. Our findings suggest that the current trends of empathy tool implementations have a limited impact on users, and more research is needed to establish good practices for eliciting long-term empathy. We discuss the current opportunities in the research conducted to date on empathic virtual environments, coined as \textit{Empathic Reality} (ER). Under the domain across XR technology, empathy, and spatiality, our survey article ends by presenting a research roadmap and several open 
research 
questions in the wide scope of digital media and the metaverse. 



\bibliographystyle{ACM-Reference-Format}
\bibliography{empathySurvey}

\appendix
\clearpage
\section{Search strings and results}
\label{appendix:search_strings}
\begin{table}[ht!]
    \centering
    \small
    \begin{tabular}{l|p{8cm}|l|l}
        \textbf{Database} & \textbf{Search string} & \textbf{Results} & \textbf{Date of search} \\
        \hline
        ACM Digital Library & Title:(("empathy" OR "sympathy" OR "pity" OR "compassion") AND ("urban" OR "space" OR "place" OR "built environment" OR "building" OR "city" OR "environment") AND ("virtual" OR "mixed" OR "augmented" OR "diminished") AND "reality") OR Keyword:(("empathy" OR "sympathy" OR "pity" OR "compassion") AND ("urban" OR "space" OR "place" OR "built environment" OR "building" OR "city" OR "environment") AND ("virtual" OR "mixed" OR "augmented" OR "diminished") AND "reality") OR Abstract:(("empathy" OR "sympathy" OR "pity" OR "compassion") AND ("urban" OR "space" OR "place" OR "built environment" OR "building" OR "city" OR "environment") AND ("virtual" OR "mixed" OR "augmented" OR "diminished") AND "reality") & 12 & 9.6.2021 \\
        \hline
        IEEE Xplore Digital Library & ("empathy" OR "sympathy" OR "pity" OR "compassion") AND ("urban" OR "space" OR "place" OR "building" OR "city" OR "environment") AND ("virtual" OR "mixed" OR "augmented" OR "diminished") AND "reality" & 39 & 8.6.2021 \\
        \hline
        Web of Science & ALL=(("empathy" OR "sympathy" OR "pity" OR "compassion") AND ("urban" OR "space" OR "place" OR "building" OR "city" OR "environment") AND ("virtual" OR "mixed" OR "augmented" OR "diminished") AND "reality") & 160 & 8.6.2021 \\
        \hline
        Scopus & TITLE-ABS-KEY=(("empathy" OR "sympathy" OR "pity" OR "compassion") AND ("urban" OR "public space" OR "social space" OR "built environment" OR "building" OR "city" OR "environment") AND ("virtual" OR "mixed" OR "augmented" OR "diminished") AND "reality") & 238 & 8.6.2021 \\
        \hline
        EBSCOHost & ("empathy" OR "sympathy" OR "pity" OR "compassion") AND ("urban" OR "space" OR "place" OR "building" OR "city" OR "environment") AND ("virtual" OR "mixed" OR "augmented" OR "diminished") AND "reality") & 171 & 8.6.2021 \\
        \hline
        Elsevier (ScienceDirect) & TITLE-ABS-KEY=(("empathy" OR "sympathy") AND ("urban" OR "place" OR "city" OR "environment") AND ("virtual" OR "augmented") AND "reality") & 12 & 9.6.2021 \\
        \hline
    \end{tabular}
    \label{tab:search_results}
\end{table}

\renewcommand{\arraystretch}{1.3}

\section{Included Publications}
\label{appendix:included_publications}
\begin{spacing}{1}
\begin{center}
\small
\begin{longtabu} to \textwidth {lp{0.7\textwidth}l}
  \hline
  Study ID & Title & Publication type \\ 
  \hline
  \endhead
  
  \hline
  \endfoot

  \hline
  225 & “Empathy Machine”: How Virtual Reality Affects Human Rights Attitudes \cite{bujicEmpathyMachineHow2020a} & Journal article \\ 
  446 & “Our Dog Probably Thinks Christmas Is Really Boring”: Re-Mediating Science Education For Feminist-Inspired Inquiry \cite{kellyOurDogProbably2020} & Conference paper \\ 
  243 & A Breathtaking Journey. On The Design Of An Empathy-Arousing Mixed-Reality Game \cite{korsBreathtakingJourneyDesign2016} & Conference paper \\ 
  201 & A Pilot Study Of The Impact Of Virtually Embodying A Patient With A Terminal Illness \cite{elziePilotStudyImpact2021} & Journal article \\ 
  238 & A Virtual Reality Simulation Of Drug Users’ Everyday Life: The Effect Of Supported Sensorimotor Contingencies On Empathy \cite{christofiVirtualRealitySimulation2020} & Journal article \\ 
   12 & A Vr Game To Teach Underwater Sustainability While Diving \cite{l.calviVRGameTeach2017} & Conference paper \\ 
  214 & Alzheimer’s Eyes Challenge: The Gamification Of Empathy Machines \cite{kromaAlzheimerEyesChallenge2018} & Conference paper \\ 
   19 & An Immersive Approach To Visualizing Perceptual Disturbances \cite{g.m.rodriguezImmersiveApproachVisualizing2017} & Conference paper \\ 
  209 & An Inquiry-Based Augmented Reality Mobile Learning Approach To Fostering Primary School Students’ Historical Reasoning In Non-Formal Settings \cite{efstathiouInquirybasedAugmentedReality2018} & Journal article \\ 
  166 & Augmenting Museum Communication Services To Create Young Audiences \cite{nechitaAugmentingMuseumCommunication2019} & Journal article \\ 
  170 & Bio-Adaptive Social Vr To Evoke Affective Interdependence - Dynecom \cite{salminenBioadaptiveSocialVR2018} & Conference paper \\ 
  198 & Building Long-Term Empathy: A Large-Scale Comparison Of Traditional And Virtual Reality Perspective-Taking \cite{herreraBuildingLongtermEmpathy2018} & Journal article \\ 
  268 & Can Virtual Reality Simulations Be Used As A Research Tool To Study Empathy, Problems Solving And Perspective Taking Of Educators? Theory, Method And Application \cite{tettegahCanVirtualReality2006} & Conference paper \\ 
  463 & Communicating Sustainable Consumption And Production In 360° Video \cite{munteanCommunicatingSustainableConsumption2020} & Conference paper \\ 
  187 & Conceptual Knowledge And Sensitization On Asperger's Syndrome Based On The Constructivist Approach Through Virtual Reality \cite{hadjipanayiConceptualKnowledgeSensitization2020} & Journal article \\ 
   13 & Cyberball3d+: A 3d Serious Game For fMRI Investigating Social Exclusion And Empathy \cite{e.mavromihelakiCyberball3D3DSerious2014} & Conference paper \\ 
   22 & Degraded Reality: Using Vr/Ar To Simulate Visual Impairments \cite{p.r.jonesDegradedRealityUsing2018} & Conference paper \\ 
    6 & Design Of Visual Deficit Simulation For Integration Into A Geriatric Physical Diagnosis Course \cite{d.alexanderDesignVisualDeficit2020} & Conference paper \\ 
  386 & Designing A Virtual Environment For Teacher Training: Enhancing Presence And Empathy \cite{stavrouliaDesigningVirtualEnvironment2018} & Conference paper \\ 
  189 & Do Medical Students Respond Empathetically To A Virtual Patient? \cite{deladismaMedicalStudentsRespond2007} & Journal article \\ 
  444 & E-Mpathy And The Phantom Limb Sensation: A Multisensory Experience For Embodiment Of Amputation \cite{guareseEmpathyPhantomLimb2020} & Conference paper \\ 
  338 & Educating Bicycle Safety And Fostering Empathy For Cyclists With An Affordable And Game-Based Vr App \cite{wangEducatingBicycleSafety2016} & Conference paper \\ 
  506 & Effectiveness Of A Poverty Simulation In Second Life®: Changing Nursing Student Attitudes Toward Poor People \cite{menzelEffectivenessPovertySimulation2014} & Journal article \\ 
  431 & Effectiveness Of Using An Immersive And Interactive Virtual Reality Learning Environment To Empower Students In Strengthening Empathy And Mastery Learning \cite{abadiaEffectivenessUsingImmersive2019} & Conference paper \\ 
  127 & Embodying Senescence: Performing Agedness In The Space Of Virtuality \cite{tanEmbodyingSenescencePerforming2019} & Journal article \\ 
  355 & Empathic Mixed Reality: Sharing What You Feel And Interacting With What You See \cite{piumsomboonEmpathicMixedReality2017} & Conference paper \\ 
  334 & Empathizing Audiovisual Sense Impairments: Interactive Real-Time Illustration Of Diminished Sense Perception \cite{werfelEmpathizingAudiovisualSense2016} & Conference paper \\ 
  245 & Empathy And Embodied Experience In Virtual Environment: To What Extent Can Virtual Reality Stimulate Empathy And Embodied Experience? \cite{shinEmpathyEmbodiedExperience2018} & Journal article \\ 
  410 & Empathy Tool Design-Eye Disease Simulator Based On Mixed-Reality Technology \cite{baiEmpathyToolDesignEye2019} & Conference paper \\ 
  182 & Empathy Toward Virtual Humans Depicting A Known Or Unknown Person Expressing Pain \cite{bouchardEmpathyVirtualHumans2013} & Journal article \\ 
  175 & Exploring Virtual Reality Experiences Of Slum Tourism \cite{griffinExploringVirtualReality} & Journal article \\ 
  405 & Feel Autism Vr – Adding Tactile Feedback To A Vr Experience \cite{haddodFeelAutismVR2019} & Conference paper \\ 
  612 & Gotta Catch ‘Em All: Exploring The Use Of Pokémon Go To Enhance Cognition And Affect \cite{allowayGottaCatchEm2021} & Journal article \\ 
   26 & Have Experience: An Investigation Into Vr Empathy For Panic Disorder \cite{v.russellHAVEExperienceInvestigation2018} & Conference paper \\ 
  637 & Hearmevirtual Reality: Using Virtual Reality To Facilitate Empathy Between Hearing Impaired Children And Their Parents \cite{embolHearMeVirtualRealityUsing} & Journal article \\ 
  190 & How Does It Feel To Be A Woman Victim Of Sexual Harassment? The Effect Of 360°-Video-Based Virtual Reality On Empathy And Related Variables \cite{venturaHowDoesIt2021} & Journal article \\ 
  219 & Impact Of Immersing University And High School Students In Educational Linear Narratives Using Virtual Reality Technology \cite{calvertImpactImmersingUniversity2020} & Journal article \\ 
  161 & Inclusion Of Third-Person Perspective In Cave-Like Immersive 3d Virtual Reality Role-Playing Games For Social Reciprocity Training Of Children With An Autism Spectrum Disorder \cite{tsaiInclusionThirdpersonPerspective2021} & Journal article \\ 
  552 & Increasing Awareness, Sensitivity, And Empathy For Alzheimer’s Dementia Patients Using Simulation \cite{campbellIncreasingAwarenessSensitivity2021} & Journal article \\ 
  629 & Initial Evaluation Of Accessibility And Design Awareness With 3-D Immersive Environments \cite{rosascoInitialEvaluationAccessibility2019} & Journal article \\ 
  411 & Living Room Conservation: A Virtual Way To Engage Participants In Insect Conservation \cite{claybornLivingRoomConservation2019} & Journal article \\ 
  172 & Myshoes – The Future Of Experiential Dementia Training? \cite{adefilaMyShoesFutureExperiential2016} & Journal article \\ 
  167 & Neuroanatomical Basis Of Concern-Based Altruism In Virtual Environment \cite{patilNeuroanatomicalBasisConcernbased2018} & Journal article \\ 
   20 & Parkinson’s Disease Simulation In Virtual Reality For Empathy Training In Medical Education \cite{y.j.liParkinsonDiseaseSimulation2021} & Conference paper \\ 
  538 & Physical And Social Presence In Collaborative Virtual Environments: Exploring Age And Gender Differences With Respect To Empathy \cite{felnhoferPhysicalSocialPresence2014a} & Journal article \\ 
   18 & Promoting Sustainability Through Virtual Reality: A Case Study Of Climate Change Understanding With College Students \cite{m.poslusznyPromotingSustainabilityVirtual2020} & Conference paper \\ 
  180 & Reducing Risk And Improving Maternal Perspective-Taking And Empathy Using Virtual Embodiment \cite{hamilton-giachritsisReducingRiskImproving2018} & Journal article \\ 
   25 & Simulating Illness: Experiencing Visual Migraine Impairments In Virtual Reality \cite{s.misztalSimulatingIllnessExperiencing2020} & Conference paper \\ 
  547 & Simulating The Cuban Missile Crisis: Crossing Time And Space In Virtual Reality \cite{stoverSimulatingCubanMissile2007} & Journal article \\ 
  638 & Stop Catcalling - A Virtual Environment Educating Against Street Harassment \cite{helgertStopCatcallingVirtual2021} & Conference paper \\ 
  563 & The Effects Of A Vr Intervention On Career Interest, Empathy, Communication Skills, And Learning With Second-Year Medical Students \cite{washingtonEffectsVRIntervention2019} & Journal article \\ 
  289 & The Virtual Doppelganger: Effects Of A Virtual Reality Simulator On Perceptions Of Schizophrenia \cite{kalyanaramanVirtualDoppelgangerEffects2010} & Journal article \\ 
  509 & Through The Eyes Of A Bystander: Understanding Vr And Video Effectiveness On Bystander Empathy, Presence, Behavior, And Attitude In Bullying Situations \cite{mcevoyEyesBystanderUnderstanding2015} & Thesis \\ 
  204 & To Be There When It Happened: Immersive Journalism, Empathy, And Opinion On Sexual Harassment \cite{steinfeldBeThereWhen2020} & Journal article \\ 
  351 & Towards The Design Of A Virtual Heritage Experience Based On The World-As-Support Interaction Paradigm \cite{schaperDesignVirtualHeritage2017} & Conference paper \\ 
  322 & Understanding Empathy Training With Virtual Patients \cite{kleinsmithUnderstandingEmpathyTraining2015} & Journal article \\ 
  222 & Using 360-Video Virtual Reality To Influence Caregiver Emotions And Behaviors For Childhood Literacy \cite{raduUsing360VideoVirtual2021} & Journal article \\ 
  378 & Using Augmented Reality And Virtual Environments In Historic Places To Scaffold Historical Empathy \cite{sweeneyUsingAugmentedReality2018a} & Journal article \\ 
  203 & Using Virtual Reality And 360-Degree Video In The Religious Studies Classroom: An Experiment \cite{johnsonUsingVirtualReality2018} & Journal article \\ 
  169 & Virtual Reality As A Tool For Environmental Conservation And Fundraising \cite{nelsonVirtualRealityTool2020} & Journal article \\ 
  475 & Virtual Reality Clinical-Experimental Tests Of Compassion Treatment Techniques To Reduce Paranoia \cite{brownVirtualRealityClinicalexperimental2020} & Journal article \\ 
  223 & Virtual Reality Perspective-Taking Increases Cognitive Empathy For Specific Others \cite{vanloonVirtualRealityPerspectivetaking2018} & Journal article \\ 
  244 & Virtual Reality, Augmented Reality, Mixed Reality, And The Marine Conservation Movement \cite{mcmillanVirtualRealityAugmented2017} & Journal article \\ 
  171 & Virtual Superheroes: Using Superpowers In Virtual Reality To Encourage Prosocial Behavior \cite{rosenbergVirtualSuperheroesUsing2013} & Journal article \\ 
  107 & Virtually Old: Embodied Perspective Taking And The Reduction Of Ageism Under Threat \cite{ohVirtuallyOldEmbodied2016} & Journal article \\ 
  323 & Visuospatial Transformations And Personality: Evidence Of A Relationship Between Visuospatial Perspective Taking And Self-Reported Emotional Empathy \cite{sulpizioVisuospatialTransformationsPersonality2015} & Journal article \\ 
   15 & Walking In Small Shoes: Investigating The Power Of Vr On Empathising With Children’s Difficulties \cite{v.camilleriWalkingSmallShoes2017} & Conference paper \\ 
  174 & What Can Virtual Reality Teach Us About Prosocial Tendencies In Real And Virtual Environments? \cite{gillathWhatCanVirtual2008} & Journal article \\ 
  436 & Who Wants To Be A Self-Driving Car? A Mixed-Reality Data-Trust Exercise \cite{leeWhoWantsBe2019} & Journal article \\ 
   \hline
\end{longtabu}
\end{center}
\end{spacing}

\end{document}